\documentclass[transmag]{IEEEtran}
\usepackage{latexsym}
\usepackage{graphicx}
\usepackage{graphics}
\usepackage{amsfonts,amssymb,amsmath}
\usepackage{hyperref}
\usepackage{todonotes}
\usepackage{cite}
\usepackage{enumitem}
\usepackage{multirow}
\usepackage{booktabs}
\usepackage[normalem]{ulem}

\ifCLASSOPTIONcompsoc
\usepackage[caption=false,font=normalsize,labelfont=sf,textfont=sf]{subfig}
\else
\usepackage[caption=false,font=footnotesize]{subfig}
\fi

\def\BibTeX{{\rm B\kern-.05em{\sc i\kern-.025em b}\kern-.08em T\kern-.1667em\lower.7ex\hbox{E}\kern-.125emX}}
\begin{document}

\title{Phase Noise Resilient Three-Level Continuous-Phase Modulation for DFT-Spread OFDM}

\author{Markku Renfors, Ismael Peruga Nasarre, Toni Levanen, Mikko Valkama, and Kari Pajukoski
\thanks{{Manuscript received {September 24, 2021}} This work was supported by the Academy of Finland under the grants \#319994 and \#332361, by the  Finnish Funding Agency for Innovation under the project 5G-FORCE, and by Nokia Bell Labs.}
\thanks{M. Renfors, I. Peruga, and M. Valkama are with Tampere University, Finland (firstname.lastname@tuni.fi).}
\thanks{T. Levanen was earlier with Tampere University, Finland. Currently, he is with Nokia Mobile Networks, Finland (toni.a.levanen@nokia.com).}
\thanks{K. Pajukoski is with Nokia Bell Labs, Finland (kari.pajukoski@nokia-bell-labs.com).}
}

\IEEEtitleabstractindextext{\begin{abstract}In this paper, a novel OFDM-based waveform with low peak-to-average power ratio (PAPR) and high robustness against phase noise (PN) is presented. It follows the discrete Fourier transform spread orthogonal frequency division multiplexing (DFT-s-OFDM) signal model. This scheme, called 3MSK, is inspired by continuous-phase frequency shift keying (FSK), but it uses three frequencies in the baseband model -- specifically, 0 and $\pm f_{symbol}/4$, where $f_{symbol}$ is the symbol rate -- which effectively constrains the phase transitions between consecutive symbols to 0 and $\pm \pi/2$ rad. Motivated by the phase controlled model of modulation, different degrees of phase continuity can be achieved, allowing to reduce the out-of-band (OOB) emissions of the transmitted signal, while supporting receiver processing with low complexity. Furthermore, the signal characteristics are improved by generating an initial time-domain nearly constant envelope signal at higher than the symbol rate. This helps to reach smooth phase transitions between 3MSK symbols, while the information is encoded in the phase transitions. Also the possibility of using excess bandwidth is investigated by transmitting additional non-zero subcarriers outside active subcarriers of the basic DFT-s-OFDM model, which provides the capability to  greatly reduce the PAPR.   
Due to the fact that the information is encoded in the phase transitions, a receiver model that tracks the phase variations without needing reference signals is developed. To this end, it is shown that this new modulation is well-suited for non-coherent receivers, even under strong phase noise (PN) conditions, thus allowing to reduce the overhead of reference signals. Evaluations of this physical-layer modulation and waveform scheme are performed in terms of transmitter metrics such as PAPR, OOB emissions and achievable output power after the power amplifier (PA). Finally, coded radio link evaluations are also shown and provided, demonstrating that 3MSK has a similar BER performance as that of traditional QPSK, but with significantly lower PAPR, higher achievable output power, and possibility of using non-coherent receivers. 
\end{abstract}
\vspace{-3mm}
}

\maketitle

\section{Introduction}
\label{sec:intro}
Low-power wide-area networks (LPWANs) are expected to be one of the fundamental pillars of the upcoming radio technologies \cite{ChettriSurvey, MoznimMTCB5G, HoellerB5GLPWAN}, where a high diversity of applications exists, such as smart healthcare, factory automation, smart agriculture, wearable devices or vehicle to vehicle (VTV) communications, to name a few, 
each with different requirements in terms of latency, reliability and throughput. These discrepancies in requirements call for different types of solutions tailored for each necessity \cite{RazaOverview}. However, one common requirement from all the types of applications is the need of having low power consumption in the physical (PHY) layer \cite{AnnamalaiIoTPHY}, since extended battery lifetime and very wide network coverage are some of the most important parameters for this type of communication.

\subsection{State-of-the-Art}

The main goals of LPWAN are to offer long range connections with the lowest possible power consumption and cost. To achieve these goals, different approaches can be taken \cite{RazaOverview} including:
\begin{itemize}
    \item Low peak to average power ratio (PAPR) modulation: LPWAN need to offer communication distances ranging from few to tens of kilometres. Low PAPR modulations can result in a very efficient use of the power amplifier (PA), increasing the transmitted power, and consequently the coverage \cite{FDSSpaper}. 
    \item Narrow-band modulation techniques: The signal is transmitted in small bandwidth, which results in low noise power in the receiver, thus leading to a large coverage. An example of a standard that follows this approach is NB-IoT \cite{nbIoTTutorial}. Furthermore, using a narrow transmission bandwidth also allows for increased number of supported devices per unit bandwidth \cite{RazaOverview}.
    \item Spread spectrum techniques: Narrow-band signals are spread over a larger bandwidth, resulting in a noise-like signal. To detect the signal, the receiver processing gain needs to be larger than without spread spectrum. LoRa \cite{LoRa} utilizes direct sequence spread spectrum (DSSS) technique. 
    \item Duty cycling: By having the device radio frequency (RF) chain functioning less amount of time, the power consumption can be reduced. However, this comes with a reduced throughput and increased latency \cite{DemikolMAC}.
    \item Reduction in control signaling overhead: Similarly to duty cycling, reduction of control signaling overhead results in lower power consumption at the expenses of increased latency.
    \item Reduction in hardware complexity: Since the devices need to be low cost, they need to be able to process less complex waveforms.
\end{itemize}

From the previous listed approaches, in this work, a low PAPR modulation is proposed in order to use the PA efficiently and increase the coverage of the network, while not compromising the spectral efficiency. 

The presented 3MSK modulation is inspired by the minimum shift keying (MSK) idea, which was developed in \cite{MSKPatent}. MSK is a special form of binary continuous phase frequency shift keying (CPFSK) and continuous phase modulation (CPM), with a modulation index $h=1/2$, and a pulse shape of half of a sinusoid \cite{proakis_digital_2008}. The term "minimum" in MSK comes from the fact that the two frequencies encoding the data have minimum possible frequency deviation necessary to ensure orthogonality between both binary symbols. The resultant waveform presents a constant envelope. In \cite{MSKPasupathy}, connections of MSK waveform with QPSK, offset QPSK (OQPSK) and frequency shift keying (FSK) are explained, including the constant envelope properties as well as the error rate performance in relation to binary PSK (BPSK). 
However, MSK still presents high out-of-band (OOB) emissions, which led to the idea of Gaussian MSK (GMSK) \cite{GMSK} in order to reduce the side-lobes by using a Gaussian low-pass filter prior to the modulation.

Furthermore, the need of having robustness against phase noise (PN) in LPWAN is demonstrated in \cite{PNcompIoT}, where a PN compensation method for IoT devices is proposed for the standard for low-rate wireless networks IEEE 802.15.4 \cite{IEEE802154}, where also non-coherent detection to MSK signal is discussed as a way to reduce the power consumption of the receiver. 

The most widely studied approach for low-PAPR waveforms is based on $\pi/2$ phase-rotated BPSK modulation \cite{Kim_TVT2018}, which has well-controlled phase behaviour where the  phase rotation between consecutive BPSK symbols is $\pm \pi/2$. Due to the benefits of OFDM-based multiple access, $\pi/2$-BPSK is usually considered in DFT-s-OFDM \cite{Myung_VTMAG2006} context, also known as single-carrier frequency-division multiple access (SC-FDMA), which is applied in 4G long-term evolution (LTE), NB-IoT, as well as in 5G New Radio in coverage-limited uplink scenarios. In basic form, DFT-s-OFDM signal is transmitted in minimum number of subcarriers, which is equal to the number of BPSK (or generally QAM) symbols allocated to an OFDM symbol. This corresponds to zero roll-off in traditional single-carrier transmission schemes. However, both in traditional and DFT-s-OFDM cases, the signal characteristics, primarily PAPR and OOB emissions, can be greatly improved by using excess band. Then the number of active subcarriers is higher in the DFT-s-OFDM model. In this context, frequency-domain spectrum shaping (FDSS) \cite{FDSSpaper} is commonly applied, corresponding to Nyquist pulse shaping. This means that the subcarrier samples generated by DFT are copied symmetrically over the excess bands on both sides (corresponding to up-sampling in time-domain multirate signal processing), and then the used DFT bins are weighted, e.g., by a raised cosine (RC) type function. Various schemes for optimizing the FDSS weights have been presented in the literature, see \cite{Choi_OJ_COM_2021} and references therein. There are clear indications that the RC-type filters are not optimal for FDSS-based low-PAPR schemes. 

The block structure of OFDM based waveforms introduces abrupt transitions between OFDM symbols and also between main OFDM symbol and its cyclic prefix (CP). This is particularly an issue in DFT-s-OFDM signals with well-controlled amplitude and phase behavior, such as $\pi/2$-BPSK, CPSK \cite{CPSK}, and 3MSK, which can be considered as nearly constant-envelope (CE) waveforms. Therefore, methods to guarantee smooth/continuous phase behavior between (i) CP-OFDM symbols and (ii) between main OFDM symbol and its CP have been considered in the literature for CPM-based DFT-s-OFDM and $\pi/2$-BPSK in particular, see \cite{Wylie_TCOM_2011} and references therein.

The idea of constraining the phase transitions between consecutive PSK symbols with the goal of reduced PAPR was studied in \cite{CPSK}, where a constrained PSK (CPSK) modulation was presented for DFT-s-OFDM. For CPSK, an underlying PSK constellation is defined, and for every transmitted symbol, only a smaller set of the constellation symbols are available for transmission. 
3MSK can also be seen as a special case of CPSK with an underlying constellation of QPSK and three possible symbols available for each transmitted symbol. However, in this paper we present various additional features, including oversampled signal generation, excess band utilization, as well as application and analysis of different signal phase continuity aspects.


\subsection{Contributions and Organization}

DFT-s-OFDM based low-PAPR schemes have focused on binary modulation because it is difficult/impossible to reach the targeted PAPR and OOB characteristics with low/modest complexity with four or more modulation levels. Using 3-level modulation is unconventional, but it can offer good tradeoffs for the signal characteristics while supporting 50\% higher spectrum efficiency than binary schemes. Maximizing the bit-rate with 3-level modulation becomes complicated in terms of bit-to-symbol mapping, but transmitting 3 bits in 2 symbols is straightforward and fairly close to the maximum rate, so it is followed also in this paper. It brings some redundancy in bit mapping which can be used to enhance link performance, basically through high-rate trellis coding and relatively simple sequence detection on the receiver side.

In this paper, we propose a novel modulation scheme for OFDM-based single carrier transmission, named as 3MSK, which allows to control the PAPR and OOB emissions by means of different parameterization alternatives concerning oversampled signal generation, imposed phase continuity, as well as the possibility of transmissions using excess bandwidth (EBW). Furthermore, thanks to the phase transition model of 3MSK, we present a receiver model capable of tracking and compensating PN effects without the need of extra reference signals, thus reducing the reference signal overhead. Finally, we also demonstrate that the proposed modulation can be detected with a non-coherent receiver.

\begin{figure*}[!t]
	\centering
	\vspace{-5mm}
	\includegraphics[width=0.85\textwidth]{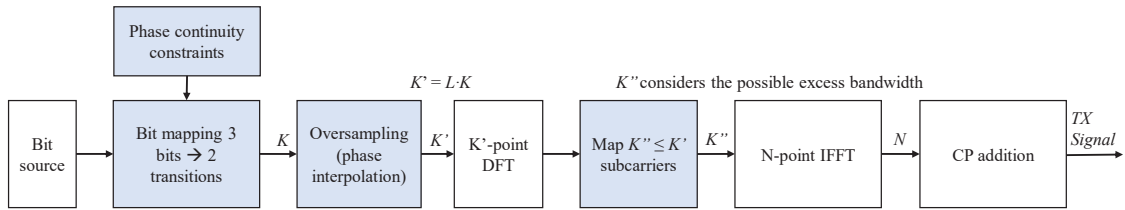}
	\vspace{-2mm}
	\caption{Block diagram of the generic 3MSK transmitter for DFT-s-OFDM.}
	\vspace{0mm}
	\label{fig:transmitterBlock}
\end{figure*} 

The novel contributions of this paper include:
\begin{itemize}
    \item 3-level based continuous phase modulation scheme with nearly constant envelope is proposed.
    \item Analysis of trellis-based bit-to-symbol mapping alternatives.
    \item Schemes for phase continuity (i) between main OFDM symbol and its CP and (ii) between CP-OFDM symbols. For (i) we adapt the model used earlier for $\pi/2$-BPSK to 3MSK. For (ii) a novel scheme based on multiples of $\pi/2$ phase rotations of CP-OFDM symbols is proposed, considering also the required receiver processing. The benefits of both elements are evaluated, leading to clear recommendations of their use.
    \item Oversampled signal generation is proposed and its capability of significant PAPR reduction is demonstrated, considering also its effects on OOB emissions and link performance. 
    \item The use of excess band together with the oversampled model is proposed and demonstrated to provide significant further reduction of PAPR. In this paper, excess band is applied only on the transmitter side and the receiver does not make use of it. This makes it possible for adjacent users/allocations to have overlapping excess bands, which reduces the related overhead up to 50 \%.
    \item Tracking of the phase error due to large PN effects without explicitly using reference signals, thus reducing the signaling overhead, and obtaining good PN compensation.
\end{itemize}

The rest of the paper is organized as follows: In Section \ref{sec:transmitter}, the 3MSK signal model and transmitter architecture are introduced, including the trellis-based bit-to-symbol mapping, phase continuity model, oversampled signal generation, and use of excess bands. Section \ref{sec:receiver} explains the receiver signal processing architecture, with focus on 3MSK signal detection. Section \ref{sec:Methods} introduces the performance metrics used for evaluation, and the numerical results and comparisons are included in Section \ref{sec:results}. Finally, the concluding remarks can be found in Section \ref{sec:conclusion}.


\section{Transmitter Processing}
\label{sec:transmitter}

3MSK is a modulation tailored for OFDM-based single carrier waveform that allows for lower PAPR, low OOB emissions, robustness against the effects of PN and can be utilized with non-coherent detection. 

\begin{figure}[!t]
    \centering
        \subfloat[]{\includegraphics[width=0.45\columnwidth]{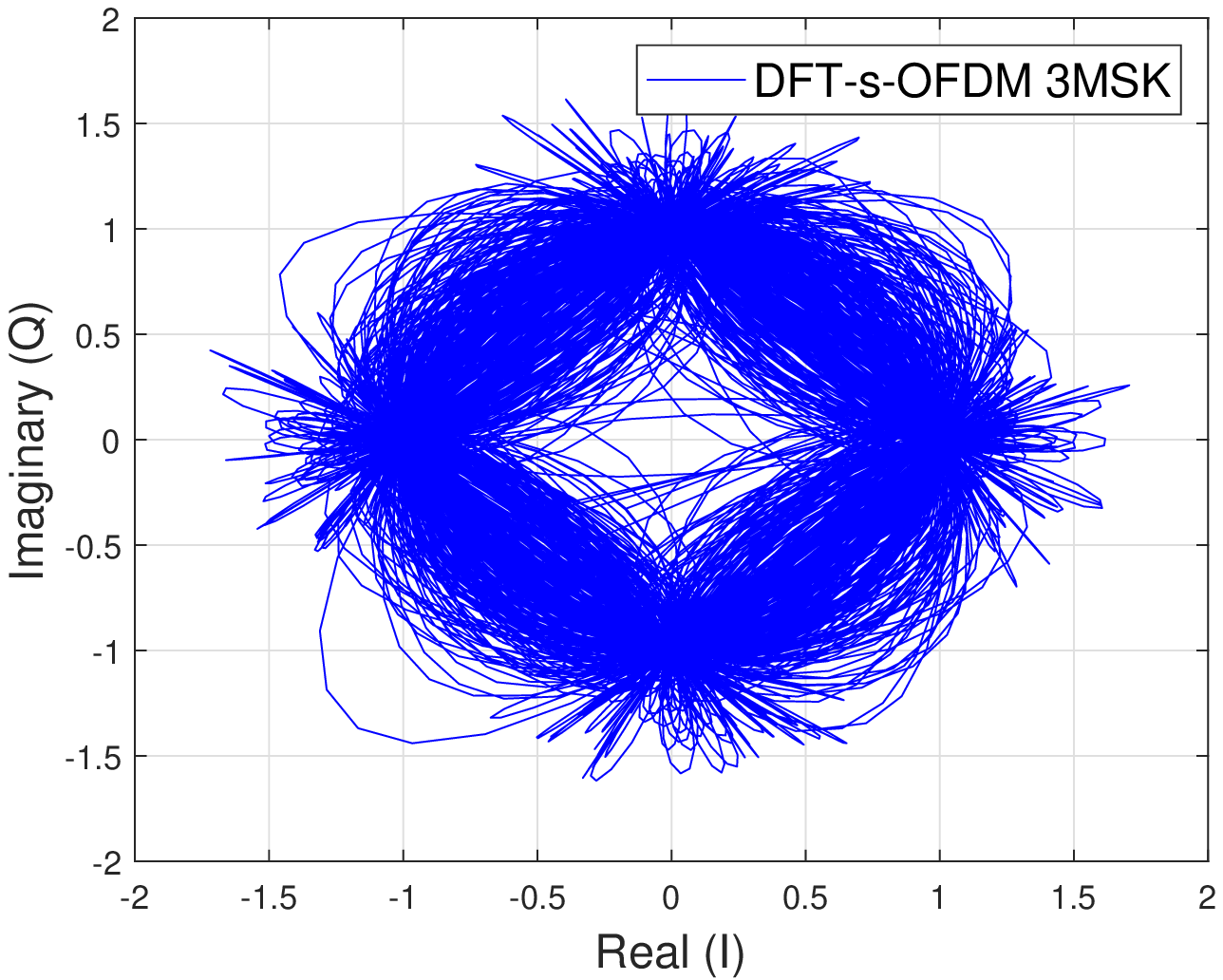}
        \label{fig:IQos1}}
        \hfil
        \subfloat[]{\includegraphics[width=0.45\columnwidth]{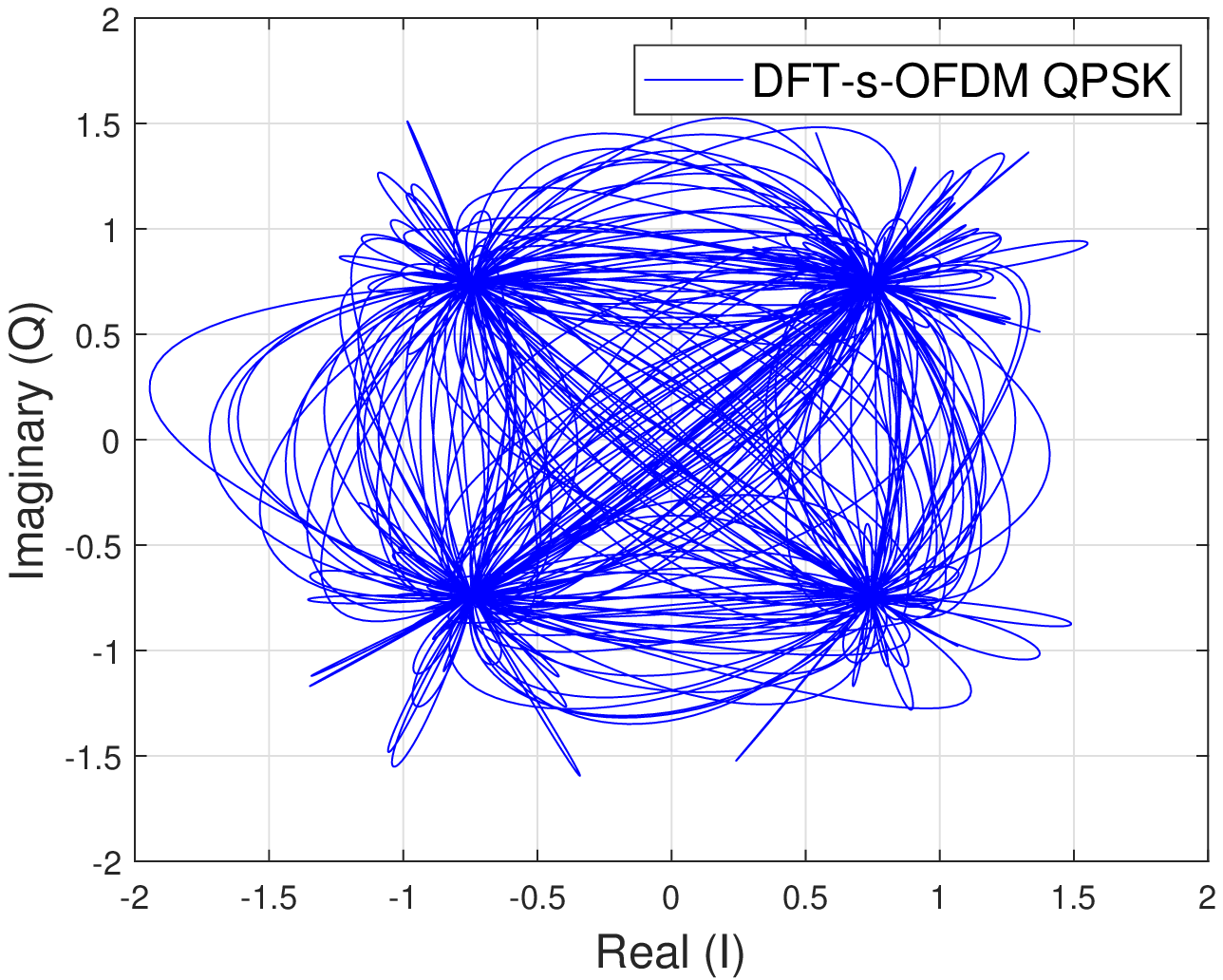}
        \label{fig:IQos2CPPC}}
    \caption{IQ scatterplot of (a) the 3MSK signal without oversampling and phase continuity options, and (b) QPSK signal for DFT-s-OFDM.}
    \label{fig:IQscatterplotsKnown}
\end{figure}

The main idea behind 3MSK relies on a continuous-phase FSK signal with three frequencies in baseband model (i.e., $0, +f_{symbol}/4$ and  $-f_{symbol}/4$, where $f_{symbol}$ is the symbol rate). The three frequencies allowed to be transmitted per symbol time effectively constrain the phase transitions between consecutive symbols to $0, +\pi/2$ and $-\pi/2$ rad. Fig. \ref{fig:transmitterBlock} shows the block diagram of the transmitter, highlighting in blue the main processing blocks for 3MSK, and Fig. \ref{fig:IQscatterplotsKnown} shows the in-phase/quadrature (IQ) scatterplots of the 3MSK signal compared to QPSK. It can be seen that by restraining the phase transitions between consecutive symbols to $\pm \pi/2$ or $0$ rad in the 3MSK modulation, the zero-crossings in the scatter-plot can be avoided, and therefore the signal does not present samples with very low power, in contrast to QPSK, where samples with low power appear in the zero-crossings, and samples with larger power than in the 3MSK are also present, increasing the PAPR of the signal.

\subsection{System Model}
Let us first define the basic quantities for presentation clarity. The mapping of bit to 3MSK symbols can be done by means of the so-called 3MSK block. A 3MSK block is a block of 3MSK symbols that follow the constraint of phase difference between consecutive symbols and encode the information in their phase transitions. $B$ bits are mapped to $K_b$ 3MSK symbols to form a 3MSK block. In the context of DFT-s-OFDM, it is natural to assign the 3MSK block size to be
the number of sub-symbols carried by a DFT-s-OFDM symbol ($K$), so that $K_{b}=K$. In traditional DFT-s-OFDM processing, the number of sub-symbols carried by a DFT-s-OFDM symbol is the same as the DFT size in the modulator, and therefore corresponds to the size of the in-band, or active subcarriers. The number of bits $B$ of a 3MSK block can be up to $3K_{b}/2$ if no other constraints than the phase difference between symbols are applied. However, as will be seen in the following sub-sections, if phase continuity between cyclic prefix (CP) and main symbol should be achieved, 2 bits less are mapped. After obtaining the $K$ 3MSK symbols, 
interpolation (sampling rate increase) by the factor of $L \geq 1$ (typically $L=2$) is performed, giving $K' = L\cdot K \geq K$ samples to be used as input to the DFT of size $K'$. 
Finally, $K''$ DFT bins are mapped to the IFFT input. An excess band of $E\in[0,N-K']$ subcarriers may be included in the mapping, such that

\begin{equation}
    K'' = K + E.
    \label{eq:Kprima2}
\end{equation}
The excess bandwidth (EBW) can also be denoted as EBW = $100\cdot E/K$ \%.


These processing steps can be interpreted as FDSS with excess band using rectangular window for weights. While this is straightforward from the implementation point of view, we have not been able to obtain significant improvement in signal characteristics or link performance using other window shapes.

\begin{figure}[]
	\centering
	\vspace{-2mm}
	\includegraphics[width=0.9\columnwidth]{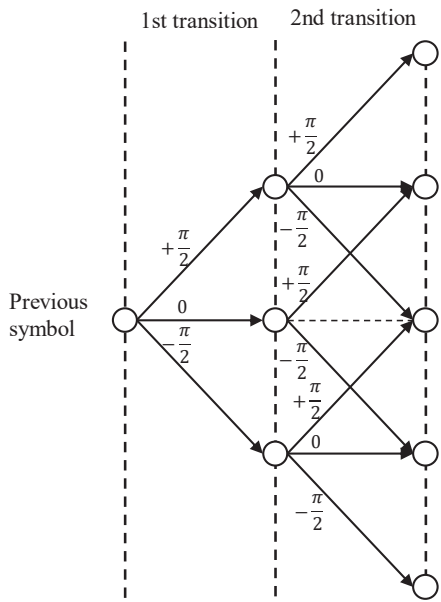}
	\vspace{-2mm}
	\caption{Example of a symmetric structure of bit-to-transition mapping. 3 bits are mapped to 8 different transitions (2 consecutive 0 rad transitions is avoided).}
	\vspace{-4mm}
	\label{fig:symmetricTransition}
\end{figure}

In the bit mapping, 3 bits are mapped into 2 consecutive transitions. This means that each 3MSK symbol carries 1.5 bits. Since 2 consecutive transition carry the 3 information bits, and the transitions can be either $0, +\pi/2, -\pi/2$, there would be 9 different transition combinations. Because 3 bits encode 8 different values, one transition combination needs to be avoided, which is illustrated in Fig. \ref{fig:symmetricTransition} showing possible transition pairs, avoiding the two consecutive 0 rad transition. Within this paper, this type of mapping is referred to as symmetric mapping (SM). It has to be noted that any of the possible transitions in the second rotation can be 
discarded, possibly generating a non-symmetric mapping (NSM). We have observed that non-symmetric mappings may provide slightly lower PAPR and OOB emissions than the symmetric mapping. However, in the continuation, we  mainly consider symmetric mapping due to its simplicity.

\begin{figure*}[!t]
    \centering
        \subfloat[]{\includegraphics[width=0.49\columnwidth]{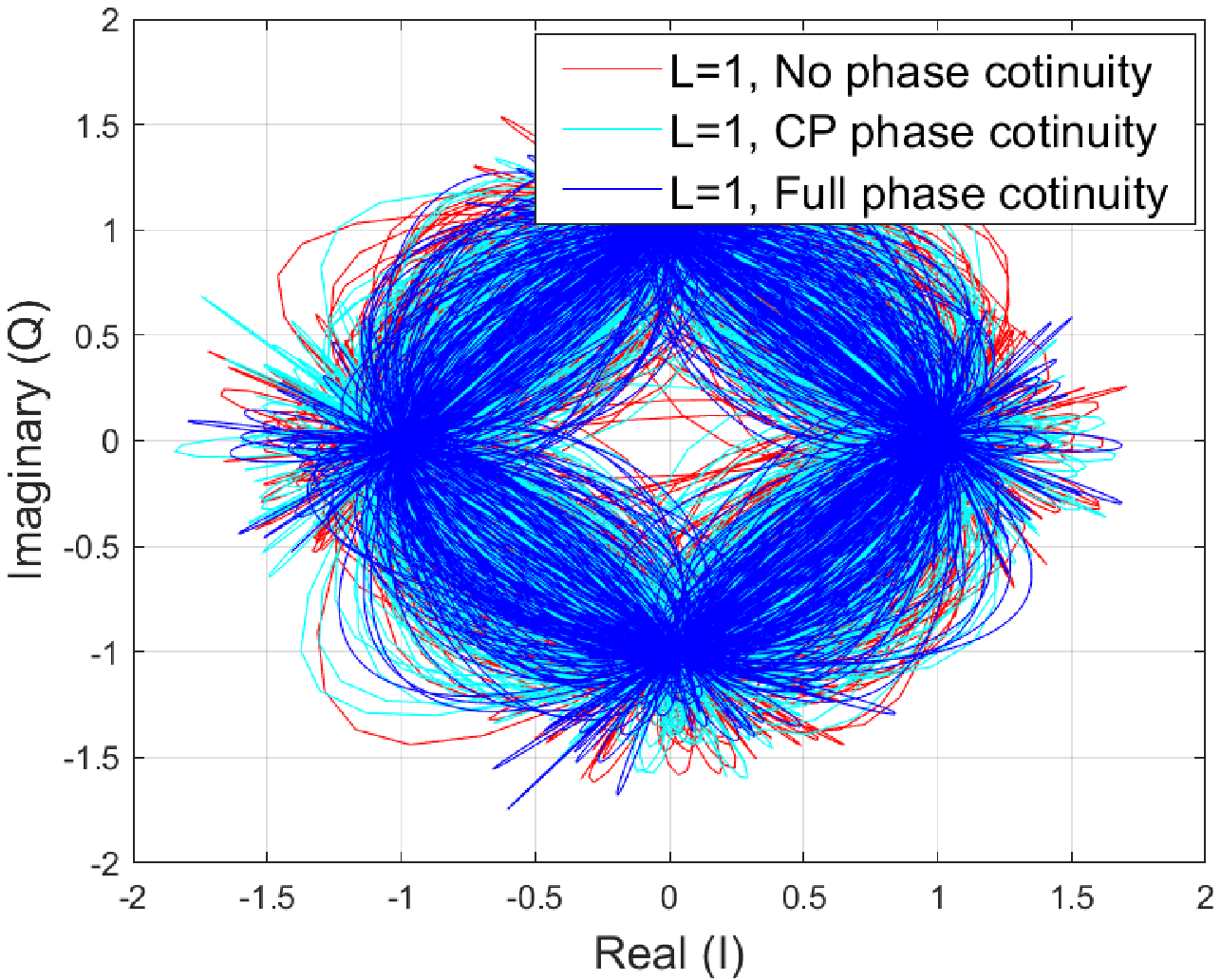}
        \label{fig:IQos1}}
        \hfil
        \subfloat[]{\includegraphics[width=0.49\columnwidth]{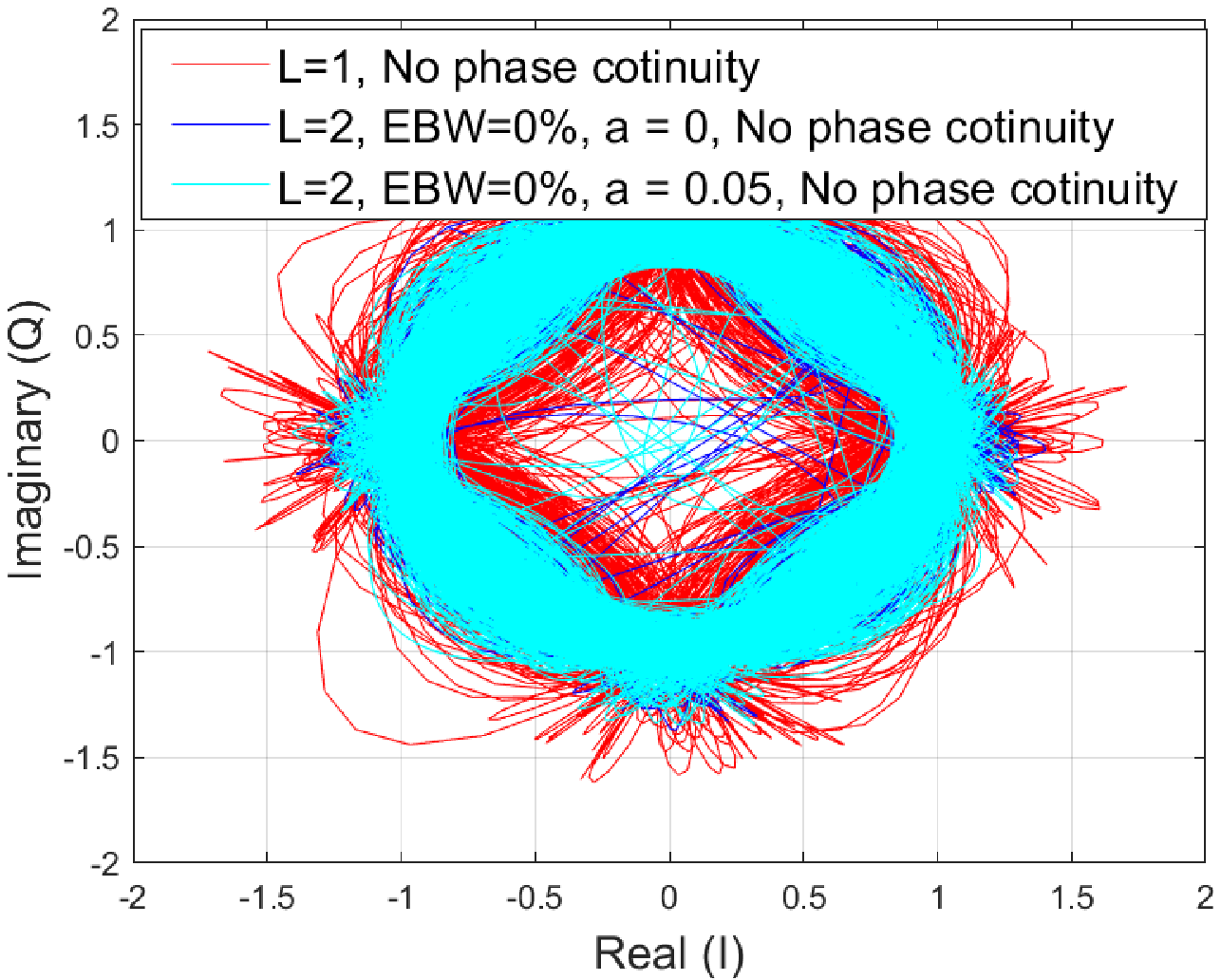}
        \label{fig:IQos1os2aComp}}
        \hfil
        \subfloat[]{\includegraphics[width=0.49\columnwidth]{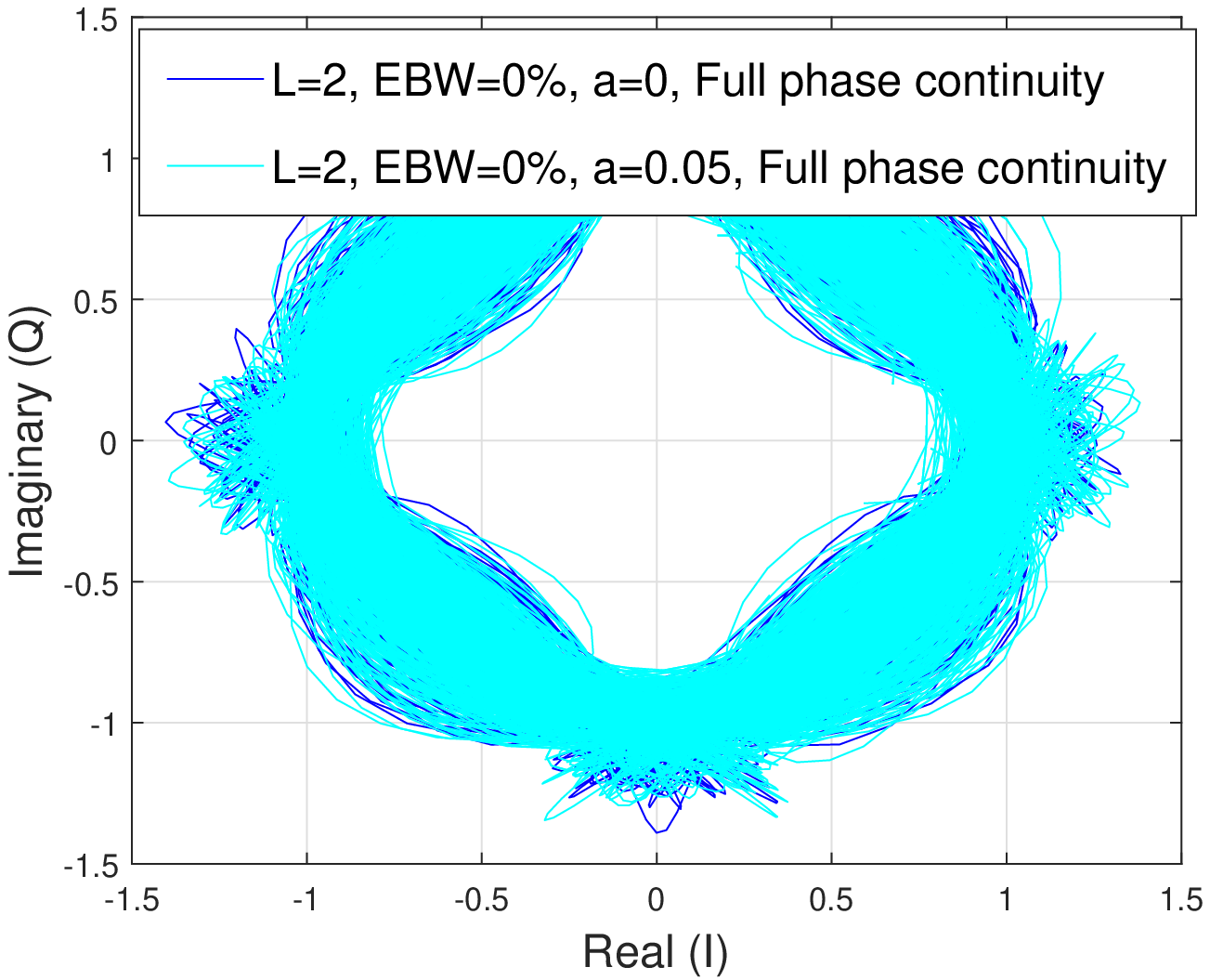}
        \label{fig:IQos2FULL}}
        \subfloat[]{\includegraphics[width=0.49\columnwidth]{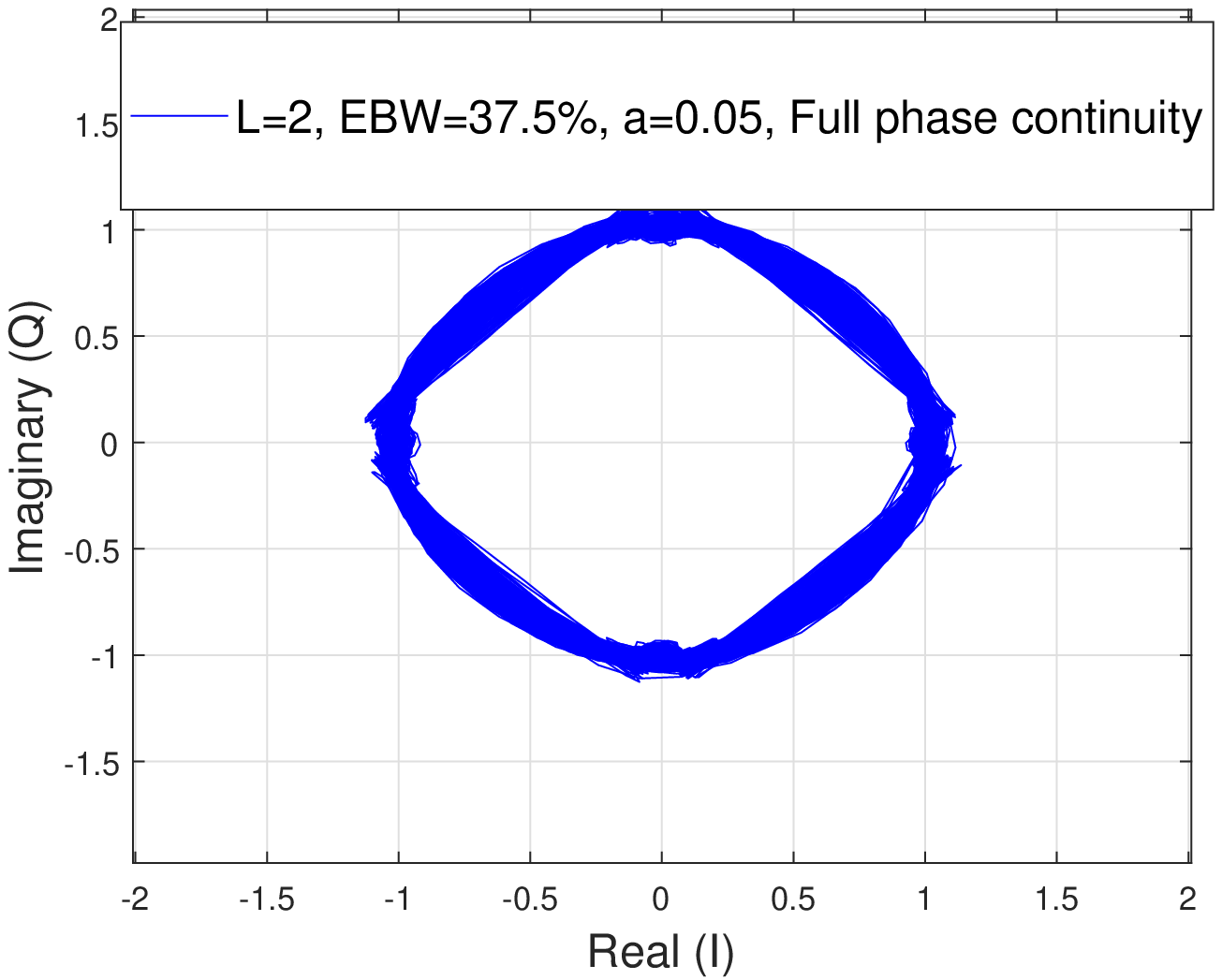}
        \label{fig:IQos2FULL20}}
    \caption{Effects of oversampling factor ($L$), excess band, and phase continuity options on the scatter plot of 3MSK. (a) $L=1$ and different phase continuities, (b) $L=1$ and $L=2$ comparison without phase continuity, (c) $L=2$ with $a=0$ and $a=0.05$ comparison with full phase continuity, and (c) $L=2$ with 37.5\% EBW.    
    }
    \label{fig:IQscatterplots}
\end{figure*} 

Table \ref{tab:bit-to-transition-Mapping} shows an example of a bit-to-transition mapping table that was found after an exhaustive search over all valid symbol sequences of 4 symbols (6 bits), based on Hamming distance and Euclidean metric for symmetric mapping. It needs to be noted that there are multiple optimum mappings that provide the same metric as the mapping of the example. Furthermore, it is clear from the nature of the mapping (bit-to-transition) that it is in fact a differential mapping, where the information is carried in the phase transition, hence allowing for non-coherent detection.

\begin{table}[]
\centering
\caption{Example of bit-to-transition mapping after exhaustive search over all valid sequences of 4 symbols.}
\resizebox{\columnwidth}{!}{%
\begin{tabular}{@{}ccccc@{}}
\toprule
b2 & b1 & b0 & \begin{tabular}[c]{@{}c@{}}First \\ transition {[}rad{]}\end{tabular} & \begin{tabular}[c]{@{}c@{}}Second \\ transition {[}rad{]}\end{tabular} \\ \midrule
0 & 0 & 0 & $-\frac{\pi}{2}$ & $+\frac{\pi}{2}$ \\
0 & 0 & 1 & $+\frac{\pi}{2}$ & $-\frac{\pi}{2}$ \\
0 & 1 & 0 & $-\frac{\pi}{2}$ & $0$ \\
0 & 1 & 1 & 0 & $-\frac{\pi}{2}$ \\
1 & 0 & 0 & 0 & $+\frac{\pi}{2}$ \\
1 & 0 & 1 & $+\frac{\pi}{2}$ & 0 \\
1 & 1 & 0 & $-\frac{\pi}{2}$ & $-\frac{\pi}{2}$ \\
1 & 1 & 1 & $+\frac{\pi}{2}$ & $+\frac{\pi}{2}$ \\ \bottomrule
\end{tabular}%
}

\label{tab:bit-to-transition-Mapping}
\end{table}

After summarizing the basic characteristics of the 3MSK, let us further explain the additional properties that this modulation can support in order to reduce PAPR and OOB emissions. This additional properties are: (i) phase continuity, either between CP and main symbol, or between consecutive DFT-s-OFDM symbols, denoted as full phase continuity when both types are included, and (ii) the oversampled transmission, with the possibility of employing excess bandwidth.

\subsection{Phase continuity}
The controlled phase transition model of the 3MSK allows us to have phase continuity between the CP and main symbol (i.e., within a DFT-s-OFDM symbol) as well as between consecutive CP-OFDM symbols. It is possible to use either of the two elements of phase continuity, but combining the two gives the maximum reduction of the OOB emissions, as will be seen in Section \ref{sec:results}.

\subsubsection{Phase continuity between CP and main symbol}
The phase continuity between CP and main symbol can be reached by forcing the phase of the first and last 3MSK symbol of the DFT-s-OFDM symbol to have a phase difference of less than, or equal to $\pi/2$ rad. Due to the cyclic nature of DFT processing, this is obtained for example by generating $K_b + 1$ 3MSK symbols, forcing the first and last symbols of the block to be the same, and just transmitting $K_b$ symbols (i.e., either the first or the last 3MSK symbol of the block is discarded). Fig. \ref{fig:IQos1} shows the scatterplot of the signal when phase continuity between CP and main symbol is used, compared to the case without forcing phase continuity and the full phase continuity case, that will be explained in the following section. It can be seen that when CP phase continuity is added, the number of zero crossings caused by phase variations of $\pi/2$ rad is reduced. However, there are still some phase transitions larger than $\pi/2$ rad that are caused by the phase discontinuity between CP-OFDM symbols. 

Assuming that the 3MSK block size is the same as the number of sub-symbols in a DFT-s-OFDM symbol, 2 bits of data per 3MSK block are lost. This is because to start and end in the same phase, only 1 bit, instead of 3, can be mapped to the last 2 transitions. The bit-to-transition mapping of the last bit of a 3MSK block depends on the phase state before the last pair of transitions, and there are 4 possible phases that could have been transmitted in the symbol before the last 2 transitions. These phases are $0, \pi/2, \pi$ and $-\pi/2$ rad. The last bit can be either $0$ or $1$. This leaves us with 8 possible transitions to which the last bit can be mapped, which allows us to to use a similar bit-to-transition mapping table as Table \ref{tab:bit-to-transition-Mapping}. This is a useful property of using the symmetric bit-to-transmission mapping, because if a non-symmetric mapping were to be used, it would not be possible to use the same table to obtain CP and main symbol continuity. 

By including the phase continuity constraint between CP and main symbol, the 3MSK block maps $3K_{b}/2 - 2$ bits to $K_{b}$ 3MSK symbols.
Assuming that the first and end phase state is the zero phase (0 rad), the bit-to-transition mapping table of the last bit to provide phase continuity between CP and main symbol is shown in Table \ref{tab:bit-to-transition-Mapping-CP-continuity}. 

This type of phase continuity helps in the receiver processing, since it can be assumed that the first and last state of the receiver trellis are the same, thus the selection of the maximum likelihood (ML) or maximum a posteriori (MAP) sequence by the detector is determined by the path that has the lowest metric with the same initial and end state. 

\begin{table}[]
\centering
\caption{Example of bit-to-transition mapping to obtain phase continuity between CP and main symbol.}
\label{tab:bit-to-transition-Mapping-CP-continuity}
\resizebox{\columnwidth}{!}{%
\begin{tabular}{@{}ccc@{}}
\toprule
\begin{tabular}[c]{@{}c@{}}Phase \\ transmitted \\ before last \\ bit {[}rad{]}\end{tabular} & \begin{tabular}[c]{@{}c@{}}First \\ transition {[}rad{]}\end{tabular} & \begin{tabular}[c]{@{}c@{}}Second \\ transition {[}rad{]}\end{tabular} \\ \midrule
0 & $-\frac{\pi}{2}$ & $+\frac{\pi}{2}$ \\
0 & $+\frac{\pi}{2}$ & $-\frac{\pi}{2}$ \\
$+\frac{\pi}{2}$ & $-\frac{\pi}{2}$ & $0$ \\
$+\frac{\pi}{2}$ & 0 & $-\frac{\pi}{2}$ \\
$-\frac{\pi}{2}$ & 0 & $+\frac{\pi}{2}$ \\
$-\frac{\pi}{2}$ & $+\frac{\pi}{2}$ & 0 \\
$\pi$ & $-\frac{\pi}{2}$ & $-\frac{\pi}{2}$ \\
$\pi$ & $+\frac{\pi}{2}$ & $+\frac{\pi}{2}$ \\ \bottomrule
\end{tabular}%
}
\end{table}

\subsubsection{Phase continuity between consecutive DFT-s-OFDM symbols}

Fig. \ref{fig:IQos1} shows the IQ scatterplot of the signal 
also when both phase continuity between consecutive DFT-s-OFDM symbols and between the CP and the main symbols are used (denoted as full phase continuity). It can be seen that in 
this case there are no abrupt phase transitions, while in the other two cases, phase transitions larger than $\pi/2$ between consecutive samples exist.

Phase continuity between consecutive DFT-s-OFDM symbols is achievable in two different ways. One alternative is to force the state sequence of the first sample of the CP to reach the same initial state from the beginning of the symbol. This approach is equivalent to the one used to generate phase continuity between CP and main symbol, and has the drawback that another 2 bits per DFT-s-OFDM symbol are lost.
\footnote{Similar approach has been used earlier with $\pi/2$ BPSK for both types of phase continuity \cite{Wylie_TCOM_2011}.}

\begin{figure}[t!]
	\centering
	\vspace{-5mm}
	\includegraphics[width=1.0\columnwidth]{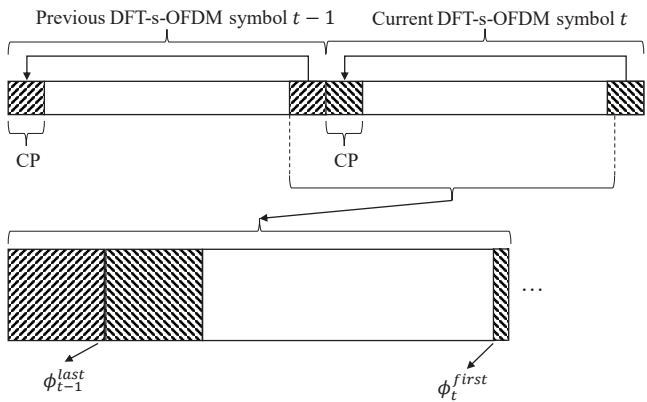}
	\vspace{-12mm}
	\caption{Phase continuity between consecutive DFT-s-OFDM symbols.}
	\vspace{-4mm}
	\label{fig:phaseContinuityDFTsOFDMSymbols}
\end{figure}

Alternatively, it is also possible to compensate the phase rotations during each 
underlaying CP-OFDM symbol in such a way that the phase continuity is achieved between the first sample of the CP and the end of the previous DFT-s-OFDM symbol, as illustrated in Fig. \ref{fig:phaseContinuityDFTsOFDMSymbols}.
In this approach, the loss of two additional bits is avoided.
Given that $\phi_{t-1}^{last}$ is the phase of the last sample of the previous ($t-1$) symbol, and the phase of the first sample of the CP of the current ($t$) symbol is $\phi_{t}^{first}$, the phase difference between these two consecutive samples is

\begin{equation}
    \phi_{t}^{diff} = \phi_{t-1}^{last} - \phi_{t}^{first}.
    \label{eq:phidiff}
\end{equation}

Then the symbol $t$ is rotated by $u \frac{\pi}{2}$ rad, where $u$ 
is the integer that satisfies

\begin{equation}
    \underset{u}{\text{min.}} \left|u \frac{\pi}{2} - \phi_{t}^{diff}\right|.
\end{equation}

It has to be noted that depending on the numerology used, when the CP-length is formed by an integer number of 3MSK symbols, the phase difference $\phi_{t}^{diff}$ will already be an integer product of $\pi/2$ rad. In the other cases when the CP-length is not formed by an integer number of 3MSK symbols, the phase continuity between DFT-s-OFDM symbols is not exact. It can be noted that the same constraints on numerology for exact phase continuity apply also for the controlled state sequence based method mentioned above.

The phase continuity between DFT-s-OFDM symbols allows to reduce the OOB emissions, and the combination with CP-main symbol phase continuity brings the largest OOB emission reduction.

\subsection{Oversampled 3MSK Signal Generation}
The 3MSK signal can be generated at higher than symbol rate, which allows for reduction of the PAPR. In this paper, we focus on the oversampling factor of $L=2$. The effects of the oversampling can be seen in Fig. \ref{fig:IQscatterplots}, where Fig. \ref{fig:IQos1} shows the non oversampled version and Fig. \ref{fig:IQos1os2aComp}, Fig. \ref{fig:IQos2FULL}, and \ref{fig:IQos2FULL20} show the scatterplots of oversampled signals. It can be seen that in the oversampled cases, the signal stays closer to the unit circle, meaning that the power variations are smaller in comparison to the non oversampled case. 

Oversampling the signal by $L=2$ is easily achievable by zero-padding between consecutive 3MSK symbols obtained after phase-to-transition mapping and applying an interpolation filter to the zero-padded sequence. 
Since the initial 3MSK signal (before DFT-s-OFDM processing) has perfectly constant envelope, a linear interpolation filter is applied to the phase of the initial 3MSK signal. This phase interpolation can be performed with any interpolation filter, but simple filters with very low complexity are found to be sufficient. One such interpolation filter can be expressed as

\begin{equation}
    h = [-a, 0, 0.5 + a, 1, 0.5 + a, 0, -a].
    \label{eq:interpolation-filter}
\end{equation}
Here $a$ is an adjustable parameter which affects the smoothness of the interpolated phase function, while the initial 3MSK phase values are not affected. With $a=0$, it corresponds to linear interpolation of phase between two consecutive 3MSK symbols.
Values of $a>0$ provide some benefits in terms of BER performance, while not significantly affecting the PAPR or OOB emissions, as will be seen in Section \ref{sec:results}. Fig. \ref{fig:IQos1os2aComp} shows the difference in the scatterplot for the signal without oversampling and the oversampled signal with different values of the parameter $a$ without imposing any phase continuity, and Fig. \ref{fig:IQos2FULL} shows the scatterplots of oversampled signal with different values of $a$ and full phase continuity. It can be seen that the effects of different $a$ values is not significant in the scatterplot, and therefore we could also expect small variations in the PAPR distribution for different $a$ values. 

To perform the interpolation, given that $x_k$, for $k \in {0, 1, ..., K_{b}-1}$ are the phases of a block 3MSK symbols of size $K_{b}$, the zero-padding between consecutive symbol phases is performed as
\begin{equation}
    \Tilde{x}_{k'} = 
    \begin{cases} 
      x_k   & \textrm{mod($k',2$)}=0 \\
      0     &  \textrm{mod($k',2$)}=1
   \end{cases}
   \label{eq:zero-padding}
\end{equation}

After the zero-padding, convolution between $h$ and $\Tilde{x}_{k'}$ gives the 2-times oversampled and interpolated phases of the symbols of the 3MSK block. It has to be noted that interpolation could also be done in the frequency domain by replicating the spectrum of the 3MSK symbols and applying the frequency-domain shape of the response of the interpolation filter $h$.

As will be seen in Section \ref{sec:results}, generating the 3MSK symbols at a higher rate helps decreasing the PAPR of the signal and affects the OOB emissions. After the oversampling and interpolation, the length of the 3MSK block is now $L\cdot K$ and a $L\cdot K$-size DFT is then performed before the subcarrier mapping.

\begin{figure}[!t]
    \centering
        \subfloat{\includegraphics[width=0.97\columnwidth]{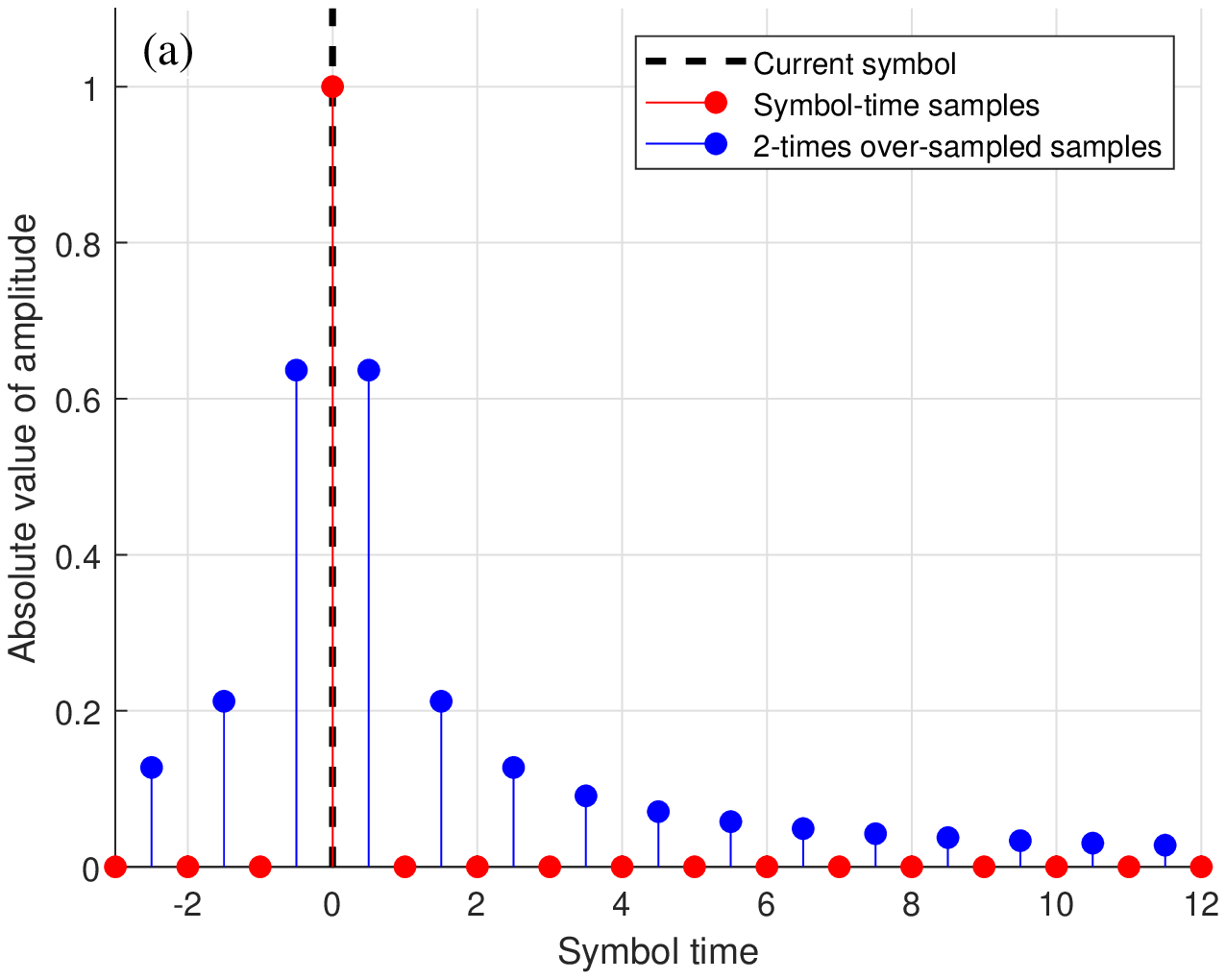}
        \label{fig:oversampled_abs}}
        \hfil
        \subfloat{\includegraphics[width=0.97\columnwidth]{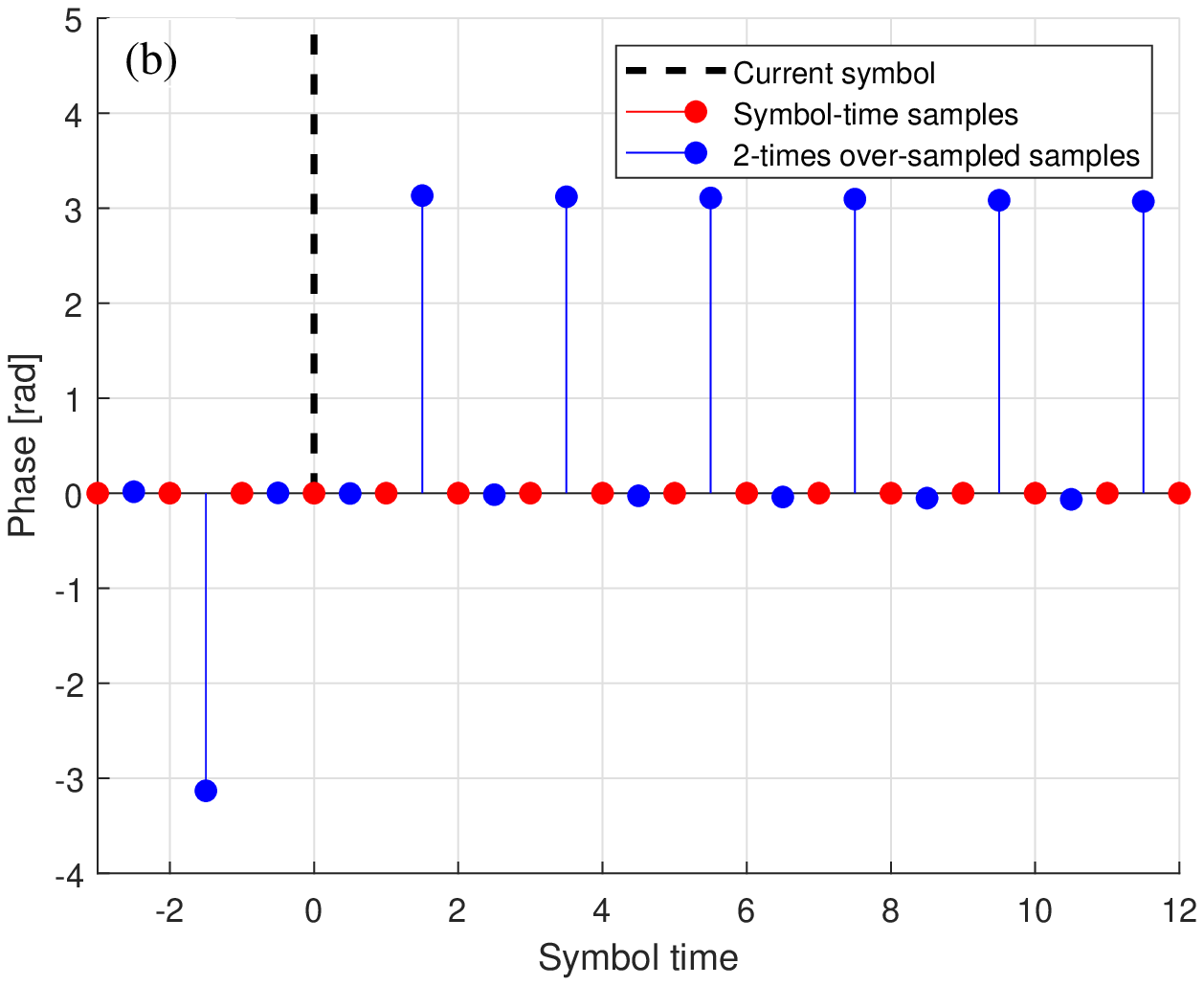}
        \label{fig:oversampled_phase}}
    \vspace{-2mm}
    \caption{Illustration of the effects of 2-times oversampling on the impulse response of DFT-s-OFDM for (a) the amplitude response and (b) phase response.} 
    \label{fig:oversampled_ISI}
\end{figure}

Fig. \ref{fig:oversampled_ISI} shows the impulse response of DFT-s-OFDM (from the DFT input to IDFT output) when 2-times oversampling is used with EBW=0 \%. It can be seen that at the original symbol-time instants (red samples), the transmission is inter-symbol interference (ISI) free since none of the symbols except the current one have impact, while the inclusion of the interpolated samples (blue samples) generate ISI, with the value of their impact decreasing at both sides of the current symbol. The inclusion of these blue samples helps to lower the PAPR, but deteriorates the receiver performance due to the added ISI. 

\subsubsection{Usage of excess band}
\begin{figure}[t!]
	\centering
	\vspace{-5mm}
	\includegraphics[width=1.0\columnwidth]{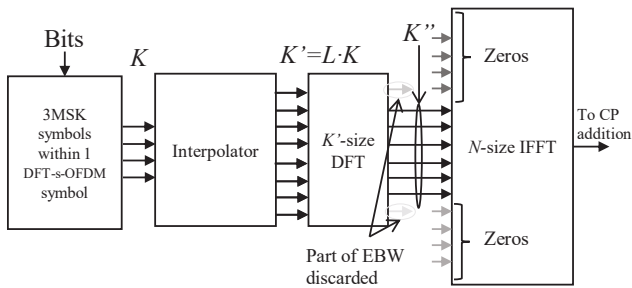}
	\vspace{-7mm}
	\caption{Diagram example of oversampled transmission with excess band and how optionally some of the bins in the excess band can be set to zero.}
	\vspace{-4mm}
	\label{fig:excessBandExample}
\end{figure}

Due to the fact that a larger DFT size is performed, more than $K$ frequency bins (also referred as subcarriers) can be used for the transmission. This allows us to use excess band, where at least a portion of the frequency bins apart from the in-band $K$ frequency bins can be allocated in the transmission, while the rest of the frequency bins remain empty. Fig. \ref{fig:excessBandExample} shows an example of how the frequency bins are obtained and allocated after the oversampling and interpolation are performed for different excess band configurations. Fig. \ref{fig:IQos2FULL20} shows the effects of using 37.5\% of excess bandwidth (EBW) in the IQ scatterplot of the signal (the percentage of EBW is computed as the number of extra active subcarriers with respect to the in-band subcarriers, as in \eqref{eq:Kprima2}). It can be seen that by using EBW, the signal envelope is even more confined compared to the case where no EBW is used, lowering the PAPR. 

The usage of excess band for transmissions helps to further decrease the PAPR of the signal. Excess band can potentially be used also in the receiver to improve the link performance, however this remains as a topic for further studies. In this case the spectral efficiency reduces further, because the excess bands of adjacent users/allocations can be partly overlapping, if the receivers do not utilize the excess band. However, if the excess band does not include all the $L\cdot K$ bins (i.e., the case when $EBW=100\%$), both in transmitter and receiver, inter-sub-symbol interference is observed in the receiver.

\section{Receiver Processing}
\label{sec:receiver}

\begin{figure*}[!t]
	\centering
	\vspace{-5mm}
	\includegraphics[width=0.85\textwidth]{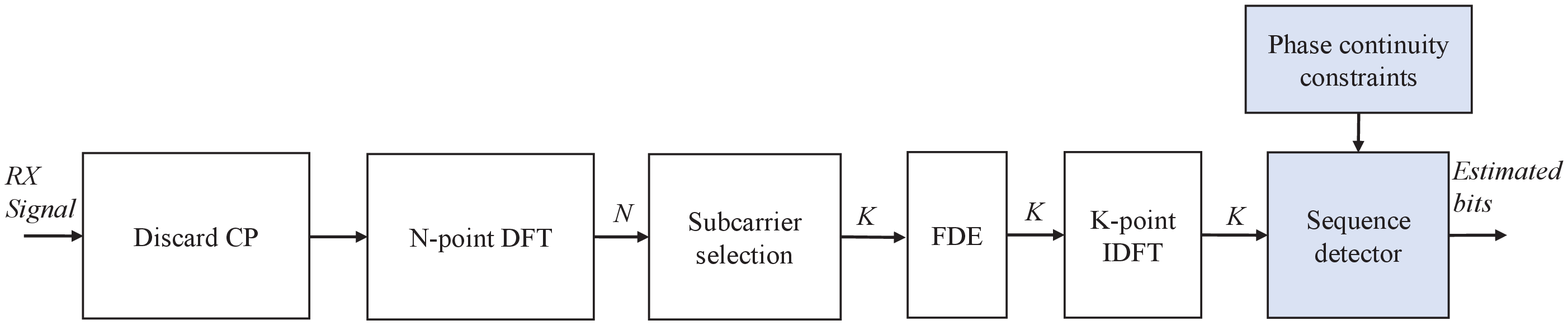}
	\vspace{-2mm}
	\caption{Block-diagram of the generic 3MSK receiver for DFT-s-OFDM. Note that even if EBW is used, the receiver only uses the in-band $K$ subcarriers for detection.}
	\vspace{0mm}
	\label{fig:receiverBlock}
\end{figure*} 

Assuming DFT-s-OFDM transmission as Fig. \ref{fig:transmitterBlock}, the receiver processing can be implemented as depicted in Fig. \ref{fig:receiverBlock}, where the blocks highlighted in blue are the additional blocks needed for 3MSK detection.

Different sequence detectors can be used to detect the 3MSK signal. In this work we consider primarily the 
BCJR algorithm \cite{BCJR}. However it has to be noted, that other types of sequence detectors can be applied as well, as the Viterbi algorithm \cite{viterbi} or soft-output Viterbi algorithm (SOVA) \cite{sova}. Actually, a modified Viterbi algorithm is used for the uncoded link performance results in Section \ref{sec:results}. In the BCJR algorithm used here, states are represented by the 3MSK symbol phases, for instance, $0$, $\pm \pi/2$ and $\pi$ rad. One additional functionality that can be added to the detection procedure is a phase error tracking estimate that updates the phase error recursively for each surviving path during the trellis search. This is very useful for mmWave and sub-THz communications, where the PN is very strong and can deteriorate the signal reception quality, especially with low-cost devices \cite{RezaiPN}. The important characteristic of this receiver processing is that it tracks the phase error by only using data symbols, i.e., no extra reference signals are needed. The addition of the recursive phase error tracking exploits the correlation between consecutive PN samples \cite{MaffezzoniPNCorrelation}.

For each surviving path, the recursive phase error estimate is updated as

\begin{equation}
    \Delta_k^n = (1-\lambda)\cdot \Delta_{k-1}^m + \lambda\cdot(\phi_{k,\textrm{observed}}-\phi_{\textrm{3MSK}}^{n}),
\end{equation}
for $m, n \in\{0,2,...,N_{\textrm{states}}-1\}$, where $N_{\textrm{states}}$ is the number of states in the trellis, $m$ is the previous state (symbol $k-1$) in the surviving path to state $n$ on the $k$th symbol (i.e., the path with the lowest metric of all the paths arriving to the state $n$ at instant $k$), $\phi_{k,\textrm{observed}}$ is the phase or the $k$th received symbol and $\phi_{\textrm{3MSK}}^{n}$ is the phase of the reference 3MSK symbol for state $n$. The constant $\lambda \in [0,1]$ is the estimation step that controls the recursive update of the phase error, giving more or less weight to the new observation. This phase error estimation can be used to compensate the received samples and improve the detection performance under severe PN degradation.

\begin{figure}[t!]
	\centering
	\vspace{-5mm}
	\includegraphics[width=1.0\columnwidth]{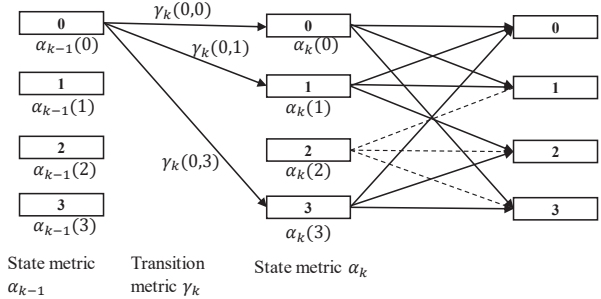}
	\vspace{-7mm}
	\caption{Example of a simple trellis diagram for the BCJR receiver for 3MSK.}
	\vspace{-4mm}
	\label{fig:trellis4}
\end{figure}

A simple trellis diagram of the receiver with $N_{\textrm{states}}=4$ is shown in Fig. \ref{fig:trellis4}. For simplicity, only the forward probabilities ($\alpha$) are shown, but the backward probabilities ($\beta$) can be computed equivalently from the opposite direction. Each state represents a complex-valued 3MSK symbol. Since it is assumed that in the transmitter the bit-to-transition mapping starts in state 0 ($s_0$, which equals symbol value of 1+0j), the only possible states for the first received symbol are $s_0$, $s_1$ and $s_2$. After receiving the first symbol, there are 3 transitions from each state (which are 0 or $\pm \pi/2)$. The trellis diagram continues until the end of the 3MSK block. The $\alpha_k(n)$ values represent the probability of being in state $n$ at time instant $k$. The updated state value follows the expression

\begin{equation}
\begin{split}
    \alpha_k(n) = \textrm{min}\{\alpha_{k-1}(m)\cdot\gamma_k(m,n)\}
    \\m,n\in\{0,2,...,N_{\textrm{states}}-1\},
\end{split}
\end{equation}
where $\gamma_k(m,n)$ is the transition metric between state $m$ and $n$ at time instant $k$. It can be based for example on Euclidean metric between the received symbol corrected by the conjugate of the phase error estimate, and the 3MSK symbol for state $m$ as

\begin{equation}
    \gamma_k(m,n) = \textrm{dist}\{r_k\cdot exp(-j\Delta_{k-1}^m), exp(j\phi_{3MSK}^{n})\},
\end{equation}

where dist\{$\cdot$\} represents a given distance (it can be for instance Euclidean distance or angular distance). $exp(-j\Delta_{k-1}^m)$ is the phase correction factor from state $m$ in the instant $k-1$, i.e., the last update of phase error estimate, $r_k$ is the received symbol at instant $k$, and $\phi_{\textrm{3MSK}}^{n}$ is the phase of the reference 3MSK symbol for state $n$.

Assuming that the 3MSK symbol block is of length $K$ and equal to the DFT length, the phase transition model includes $K+1$ states, but only $K$ of them are available from the receiver's FFT-IDFT process. Then the idea of equal initial and end states (needed for CP phase continuity) is useful also for the 3MSK signal detection because the selected trellis path would be the one with the lowest metric with the same initial and final state.

\section{Evaluation Methods and Performance Metrics}
\label{sec:Methods}

The performance of this modulation is evaluated from the transmitter and radio link level perspectives. In the transmitter side, the reduced PAPR of the signal with its different variants is evaluated, as well as the spectrum localization, measured in terms of OOB emissions and occupied bandwidth. The maximum transmit power is obtained by taking into account the RF requirements from 5G NR Rel-16 in the frequency range 2 (FR2) defined by 3GPP as the frequency bands between 24250  MHz  and  52600  MHz \cite{noauthor_3gpp_2019}. In addition, radio link level performance evaluations are performed with realistic channel models and with severe PN.

\subsection{Peak to Average Power Ratio}
The PAPR of a signal is a good first metric to measure how efficiently the PA can be used (i.e., how deep into saturation can the PA be driven without causing severe non-linearities that degrade the signal quality). In this work, the instantaneous PAPR of the signal is computed as the ratio of the power of each sample to the average power of the signal as

\begin{equation}
    {PAPR}(n) = \frac{|x(n)|^2}{\frac{1}{N_\textrm{tot}}\sum_{l = 0}^{N_\textrm{tot}-1}|x(l)|^2},
\end{equation}
where $x(n)$ is the value of the $n$th complex sample 
and $N_\textrm{tot}$ is the signal length in samples. In order to compute the PAPR, a signal with a large number of DFT-s-OFDM symbols is generated. After obtaining the instantaneous PAPR of each signal sample, the statistical distribution is presented with the complementary cumulative distribution function (CCDF) \cite{5628256}.

\subsection{Maximum Achievable Transmit Power}
To evaluate the actual transmitter performance in a more realistic scenario, the maximum output power achievable is obtained after transmitting the signal through a realistic PA model and measure the different RF emission limits defined for the 3GPP 5G NR standard with a Rel-16 compliant emission evaluation tool. 

More specifically, the RF emission requirements, defined in \cite{noauthor_3gpp_2019} are: (i) adjacent channel leakage ratio (ACLR) with a limit set to 31 dB, (ii) error vector magnitude (EVM), with a limit of 17.5\%, (iii) in-band emission (IBE), following the IBE mask defined in \cite{noauthor_3gpp_2019}, and (iv) occupied bandwidth (OBW), provided that 99\% of the power lies within the allocated bandwidth.

In order to asses the maximum achievable power, the input power of the PA is increased, measuring the PA output power and all the RF emission requirements until one requirement is not met.

The results are shown as a function of the output back-off (OBO), which corresponds to the difference in output power with respect to the saturation power of the PA, thus, 0 dB OBO corresponds to a transmission with a fully saturated PA.

\subsection{Link Level Evaluations}
A final metric to measure the performance of the studied modulation is by means of link level evaluations. For that purpose, the different variations of the 3MSK are transmitted through realistic channel models, and with the presence of PN. Uncoded bit error rate (BER) and coded block error rate (BLER) performance are obtained varying the signal to noise ratio (SNR). The coded performance is obtained by utilizing LDPC codes as defined in \cite{LDPCCodes}.

\subsubsection{Channel models}
The channel model used for evaluations is the TDL-E channel model with a delay spread of 50 ns defined in \cite{channelModels}.
\subsubsection{Phase noise models}
The PN models used for evaluations are defined in \cite{PNModels}, where the UE channel model is used in the transmitter side, and the base station (BS) channel model is used in the receiver side. For the evaluations where PN is included, a carrier frequency of 90 GHz is used, together with a subcarrier spacing (SCS) of 120 kHz.

\section{Numerical results}
\label{sec:results}


In the results shown in this paper, the size of a 3MSK block is the same as the DFT size, this means that each DFT-s-OFDM symbol with $K$ in-band subcarriers, carries one 3MSK block with $K_{b}=K$ 3MSK symbols. Table \ref{tab:sim-params} shows the main simulation parameters used.

\begin{table}[]
\centering
\caption{}
\label{tab:sim-params}
\resizebox{\columnwidth}{!}{%
\begin{tabular}{cc}
\hline
\textbf{Parameter} & \textbf{Value} \\ \hline
3MSK block size ($K_{b}$) & In-band subcarriers ($K$) \\
Phase continuity & \begin{tabular}[c]{@{}c@{}}No phase continuity, CP-main symbol,\\ between DFT-s-OFDM symbols, \\ and full phase continuity\end{tabular} \\
Subcarrier spacing & 120 kHz \\
Channel model & TDL-E \cite{channelModels}\\
Phase noise model & PN models from \cite{PNModels} \\ \hline
\end{tabular}%
}
\end{table}

\subsection{Transmitter Performance}

\begin{figure*}[!t]
	\centering
	\vspace{-5mm}
	\includegraphics[width=0.85\textwidth]{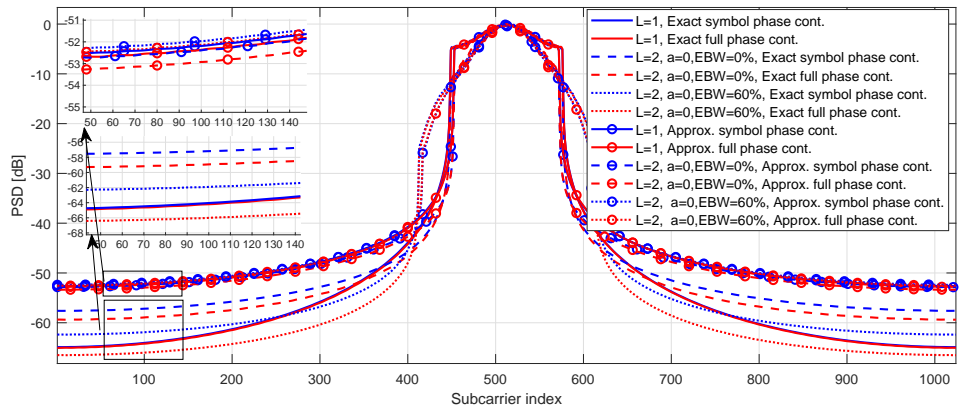}
	\vspace{-2mm}
	\caption{PSD examples of 3MSK with and without oversampling, for different combinations of phase continuity and excess bandwidth. Active subcarriers $K=128$, IFFT size $N=1024$. CP length of 128 is used to reach exact symbol phase continuity.}
	\vspace{0mm}
	\label{fig:PSDosf1osfa0}
\end{figure*} 

To evaluate the transmitter performance, let us first illustrate the effects of the different parameterization (namely phase continuity, oversampled 3MSK signal generation, EBW utilization, and symmetric/non-symmetric mapping) in the power spectral density (PSD) of the signal. Fig. \ref{fig:PSDosf1osfa0} shows the PSD of the signal with different types of phase continuity (phase continuity between CP and main symbol, phase continuity between consecutive DFT-s-OFDM symbols, and full phase continuity), as well as oversampling factors of 1 and 2, with different usage of the excess band. It is important to note that Fig. \ref{fig:PSDosf1osfa0} illustrate the effects when exact phase continuity is used between DFT-s-OFDM symbols (i.e., $\phi_{t}^{diff}$ is a multiple of $\pi/2$ in \eqref{eq:phidiff}). It is observed that 2 groups can be differentiated, based on the types of phase continuity.
\subsubsection{No phase continuity and CP phase continuity}
On one side, the group with higher OOB emissions is formed with the cases when no phase continuity is used, or when only CP phase continuity is used, irrespective of what is the oversampling factor or the usage of EBW. It can be seen that in this group, the combination providing the lowest OOB emissions corresponds to the case when oversampling is used without EBW utilization. An important observation is that when only CP to main symbol phase continuity is employed, the OOB emissions are higher than the case without phase continuity. This is due to the periodicity of the DFT/IFFT processing of DFT-s-OFDM. In the CP to main symbol phase continuity model, $K_{b} + 1$ 3MSK symbols are generated, being the first (reference) and last symbols the same, but only $K_b$ symbols are used for transmission. This means that the phase difference between the last symbol of the block, and the first transmitted symbol of the block is always lower than $\pi/2$ rad. Due to the periodicity of DFT and IFFT, the high-rate DFT-s-OFDM signal approaches the first sample value at the last sample of the symbol even when this phase continuity is imposed, i.e., some degree of phase continuity is achieved automatically. In the non-oversampled cases and oversampled cases where symbol phase continuity is not imposed, controlling the phase transition between CP and main symbol does not seem to provide benefits in terms of OOB emissions.  

\subsubsection{Symbol phase continuity and full phase continuity}
The group with the lowest OOB emissions is formed with the cases when the phase continuity between DFT-s-OFDM symbols is used, and the cases when full phase continuity is used. As expected, with full phase continuity, the OOB emissions for both oversampling factors are the lowest of all the tested cases.  However, in the 2 times oversampled case, we have more control over the phase behavior and visible effects can be seen when full phase continuity is utilized, compared to the case of just symbol phase continuity. An important observation is that with $L=2$, the PSD decays faster when going away from the in-band subcarriers, but further from the in-band $L=1$ presents lower OOB emissions.



\begin{figure*}[!t]
	\centering
	\vspace{-2mm}
	\includegraphics[width=0.85\textwidth]{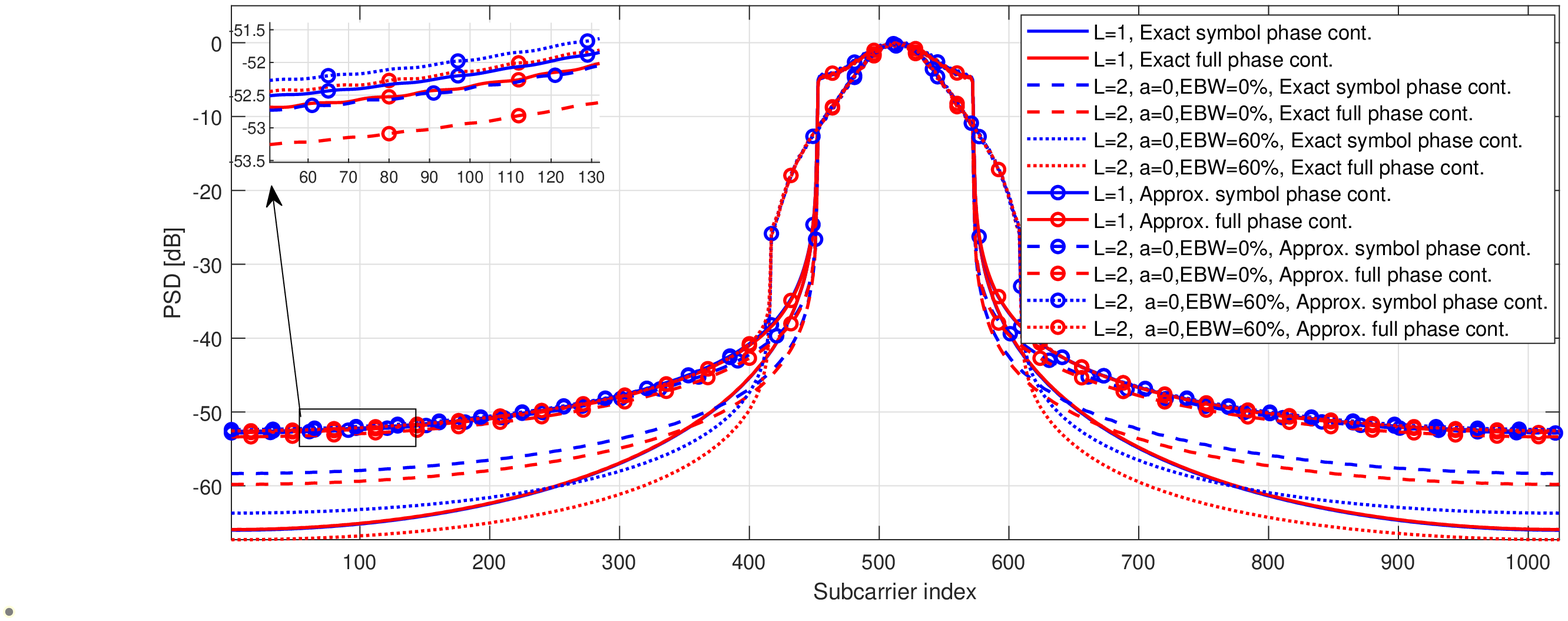}
	\vspace{-2mm}
	\caption{PSD examples of 3MSK for exact phase continuity between DFT-s-OFDM symbols and non-exact phase continuity between DFT-s-OFDM symbols. Active subcarriers $K=120$, IFFT size $N=1024$. For exact phase continuity between CP-OFDM symbols, the CP length is 128 samples, and for approximated phase continuity between CP-OFDM symbols, the CP length is 72 samples.}
	\vspace{0mm}
	\label{fig:PSD120128}
\end{figure*} 
Additionally, we study the effects of exact vs. approximate phase continuity between DFT-s-OFDM symbols. Recall that the condition for exact symbol phase continuity is that, $\phi_{t}^{diff}$ is a multiple of $\pi/2$, which is achieved when the CP-length is formed by an integer number of 3MSK symbols.
The comparison of exact and non-exact phase continuity between DFT-s-OFDM symbols is illustrated in Fig. \ref{fig:PSD120128}. It can be seen that when the phase continuity is exact, the OOB emissions can be reduced between 5 and 10 dB, compared to the case of approximate symbol phase continuity. It has to be noted that although approximated phase continuity between DFT-s-OFDM symbols is not as effective as exact phase continuity, it still reduces the OOB emissions by up to 10 dB with respect to the case where no phase continuity is forced (see Fig. \ref{fig:PSDosf1osfa0}).

\begin{figure}[t!]
	\centering
	\vspace{-5mm}
	\includegraphics[width=1.0\columnwidth]{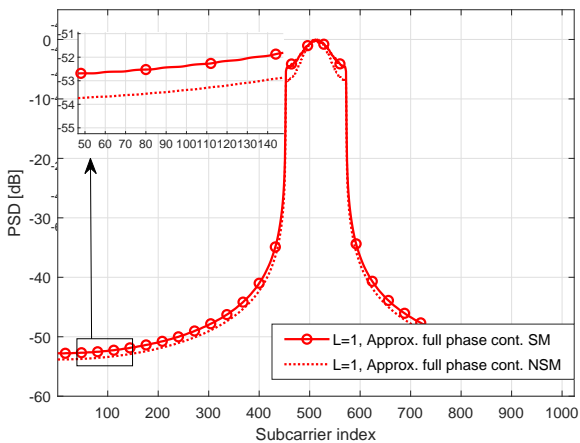}
	\vspace{-7mm}
	\caption{PSD examples of 3MSK for $L=1$ comparing  non-symmetric mapping with symmetric mapping with approximate full phase continuity.}
	\vspace{-4mm}
	\label{fig:PSD_SMvsNSM}
\end{figure}

Finally, Fig. \ref{fig:PSD_SMvsNSM} compares non-symmetric bit-to-transition mapping with symmetric mapping (SM). It can be seen that the NSM presents lower OOB emissions that the SM due to the smoother phase variations.

\begin{figure}[t!]
	\centering
	\vspace{-2mm}
	\includegraphics[width=1.0\columnwidth]{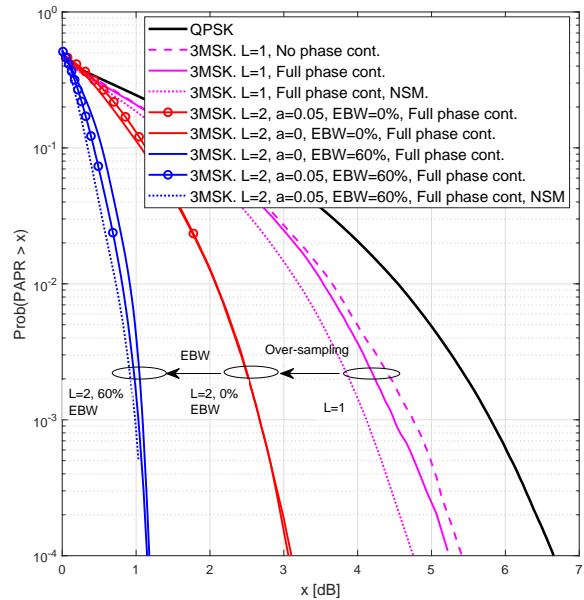}
	\vspace{-7mm}
	\caption{Example of PAPR CCDF distributions with and without oversampling, with different excess bandwidths (EBWs).}
	\vspace{-4mm}
	\label{fig:PAPR}
\end{figure}

To better characterize the effects of the different options, Fig. \ref{fig:PAPR} illustrates the comparisons of the PAPR distributions of the 3MSK signal applied on DFT-s-OFDM, compared to QPSK. It can be seen that the basic 3MSK transmission without oversampling or EBW already presents lower PAPR than QPSK (1 dB lower at $10^{-2}$ CCDF probability point). It has to be noted that if NSM is used, the PAPR can be reduced further 0.3 dB with respect to the symmetric mapping with full phase continuity. This is because NSM favors smoother phase variations when transition pair $+\pi/2,-\pi/2$ (or $-\pi/2,+\pi/2$) is discarded, instead of the transition pair $0,0$ in the symmetric mapping.

The PAPR can be further reduced with the aid of oversampling, where up to 1.5 dB lower PAPR at $10^{-2}$ CCDF probability point can be achieved if $L = 2$ and no EBW is used. Furthermore, increasing the EBW from 0\% to 60\% further reduces the PAPR down to 0.7 dB, at the expense of spectrum efficiency loss. 

Next we compare the PAPR and OOB emission performance of 3MSK against the $\pi/2$ BPSK based schemes included in the comparison of \cite{Choi_OJ_COM_2021}. Here the PAPR comparison is based on OFDM symbol based PAPR metric, instead of the sample based one used elsewhere in this paper.  The PAPR values are shown in Table \ref{tab:PAPR_schemes} at  1 \% probability level. The main parameters are N=1024, K=24 (2 physical resource blocks (PRB)), and the excess bandwidth is 50 \% when applied. The proposed 3MSK achieves clearly better PAPR performance with larger spectral efficiency when compared to the basic $\pi/2$ BPSK based schemes without excess band and FDSS with root raised cosine (RRC) weights with excess band, while advanced binary schemes presented in \cite{Choi_OJ_COM_2021} reach somewhat lower PAPR.


\begin{table}[]
\centering
\caption{PAPR comparison of the proposed 3MSK modulation and the schemes of \cite{Kuchi_ETT2012, 3GPP_low_par, Choi_OJ_COM_2021, Kim_TVT2018}}
\label{tab:PAPR_schemes}
\resizebox{\columnwidth}{!}{%
\begin{tabular}{lccc}
\hline
\textbf{Scheme}                                & \multicolumn{3}{c}{\textbf{PAPR @1\%  Probability}}         \\ \hline
                                               & \textbf{Basic scheme} & \textbf{0 EBW}  & \textbf{50\% EBW} \\ \cline{2-4} 
\textbf{3MSK $L=1$}                            & \textbf{4.9 dB}       & N/A             & N/A               \\
$\pi/2$BPSK                                    & 5.2 dB                & N/A             & N/A               \\
\textbf{3MSK $L=2$}                            & N/A                   & \textbf{2.9 dB} & \textbf{1.3 dB}   \\
Schemes of \cite{Kuchi_ETT2012} and  \cite{3GPP_low_par} & N/A                   & 1.5 - 2.1 dB    & N/A               \\
FDSS with RRC weights                          & N/A                   & N/A             & 2.2 dB            \\
Schemes of \cite{Choi_OJ_COM_2021} and \cite{Kim_TVT2018} & N/A                   & N/A             & 0.6-0.9 dB        \\ \hline
\end{tabular}%
}
\end{table}

Regarding OOB emissions, Table \ref{tab:OBW} shows the occupied bandwidths for -20 dB and -30 OOB power ratios for the same modulation schemes while assuming ideal (linear) power amplifier. We can observe that the occupied bandwidth  of 3MSK is rather similar to the best $\pi/2$ BPSK based schemes at 1\% OOB emission level, but much better at lower OOB emission levels. 
The -20 dB OOB power ratio is reached outside the band of K=24 subcarriers (normalized bandwidth of 1) both in non-oversampled and oversampled cases of 3MSK without excess band, and outside K+6=30 subcarriers (i.e., normalized bandwidth of 1.25) when 50\% excess band is applied. These values are similar to best schemes included in the comparisons of \cite{Choi_OJ_COM_2021} in terms of OOB emissions. Furthermore, -30 dB OOB power ratio is reached with normalized bandwidth of  1.42 in non-oversampled case, 1.34 in oversampled case without excess band, 1.5 with 50\% excess band, while the corresponding normalized bandwidth is beyond 2.5 in all cases included in \cite{Choi_OJ_COM_2021}.


\begin{table}[]
\centering
\caption{OOB bandwidth comparison of the proposed 3MSK modulation and the schemes of \cite{Kuchi_ETT2012, 3GPP_low_par, Choi_OJ_COM_2021, Kim_TVT2018} with linear PA}
\label{tab:OBW}
\resizebox{\columnwidth}{!}{%
\begin{tabular}{lcc}
\hline
\textbf{Scheme}               & \multicolumn{2}{c}{\textbf{Normalized bandwidth}}                           \\ \hline
                              & \textbf{-20 dB OOB} & \textbf{-30 dB OOB} \\ \cline{2-3} 
\textbf{3MSK $L=1$}           & \textbf{1}                           & \textbf{1.42}                        \\
$\pi/2$BPSK                   & 1.18                                 & \textgreater{}\textgreater{}2.5      \\
\textbf{3MSK $L=2$, 0\% EBW}  & \textbf{1}                           & \textbf{1.34}                        \\
Schemes of \cite{Kuchi_ETT2012} and  \cite{3GPP_low_par}               & 0.95-1                               & \textgreater{}\textgreater{}2.5      \\
\textbf{3MSK $L=2$, 50\% EBW} & \textbf{1.25}                        & \textbf{1.5}                         \\
FDSS with RRC                 & 1.38                                 & \textgreater{}\textgreater{}2.5      \\
Schemes of \cite{Choi_OJ_COM_2021} and \cite{Kim_TVT2018}               & 1.20-1.26                            & \textgreater{}\textgreater{}2.5      \\ \hline
\end{tabular}%
}
\end{table}

Table \ref{tab:obw} shows the occupied bandwidth in terms of normalized bandwidth for the different phase continuity and EBW configurations, with different values of the parameter $a$. A linear (ideal) PA is used to obtain the reference values, and a modified Rapp model from \cite{80211adRapp} with an input backoff (IBO) of 0.5 dB is used to evaluate the signal with non-linear PA. It can be seen that modifying the parameter $a$ of the interpolation filter does not have any effect on spectrum localization. An important result, showed in Fig. \ref{fig:PSDosf1osfa0} as well, is that full phase continuity has a significant effect in terms of occupied bandwidth.

\begin{table}[]
\centering
\caption{Occupied bandwidth in terms of normalized bandwidth for different phase continuity (PC) and EBW configurations.}
\label{tab:obw}
\resizebox{\columnwidth}{!}{%
\begin{tabular}{@{}lcccc@{}}
\toprule
                                   & \multicolumn{2}{c}{\textbf{@-20 dB}}     & \multicolumn{2}{c}{\textbf{@-30 dB}}     \\ \cmidrule(l){2-5} 
                                   & \textbf{Linear PA} & \textbf{IBO=0.5 dB} & \textbf{Linear PA} & \textbf{IBO=0.5 dB} \\ \cmidrule(l){2-5} 
$L=1$, no PC                       & 1.17               & 1.92                & $>7$               & $>7$                 \\
$L=1$, full PC                     & 1                  & 1.42                & 1.42               & 2.67                \\
$L=2$, EBW=0\%, no PC, a = 0       & 1.08               & 1.42                & $>7$                & $>7$                 \\
$L=2$, EBW=0\%, full PC, a = 0     & 1                  & 1                   & 1.33               & 2                   \\
$L=2$, EBW=0\%, no PC, a = 0.05    & 1.08               & 1.42                & $>7$               & $>7$                 \\
$L=2$, EBW=0\%, full PC, a = 0.05  & 1                  & 1.08                & 1.33               & 2.08                \\
$L=2$, EBW=50\%, no PC, a = 0      & 1.42               & 1.5                 & $>7$               & $>7$               \\
$L=2$, EBW=50\%, full PC, a = 0    & 1.25               & 1.25                & 1.5                & 2.17                \\
$L=2$, EBW=50\%, no PC, a = 0.05   & 1.42               & 1.5                 & $>7$               & $>7$                \\
$L=2$, EBW=50\%, full PC, a = 0.05 & 1.25               & 1.25                & 1.5                & 2.08                \\ \bottomrule
\end{tabular}%
}
\end{table}

\begin{figure}[t!]
	\centering
	\vspace{0mm}
	\includegraphics[width=1.0\columnwidth]{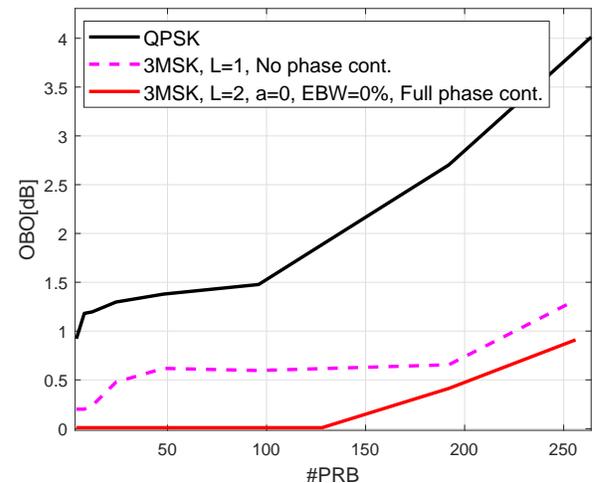}
	\vspace{-7mm}
	\caption{Simulated achievable OBO comparison between QPSK and 3MSK.}
	\vspace{-4mm}
	\label{fig:OBO}
\end{figure}

Finally, a way to evaluate the effects of the PAPR reduction on the signal is to use a realistic PA model and consider the different RF requirements that the signal needs to comply with (namely ACLR, EVM, IBE, OBE, and OBW) in order to obtain what is the maximum output power after the amplification stage. Fig. \ref{fig:OBO} shows the simulated OBO with respect to the PA saturation point, for an FR2 PA \cite{80211adRapp} for 3MSK with and without oversampling and QPSK, with respect to the number of allocated PRBs. One PRB corresponds to 12 subcarriers, in 3GPP nomenclature. It is shown that the basic 3MSK without oversampling can already output up to 3 dB higher power than QPSK when the channel is fully allocated, and 1 dB more for small allocations. Furthermore, with $L=2$ and full phase continuity the PA can be driven to full saturation for small and medium size allocations, and with full allocation, extra 0.4 dB output power compared to non-oversampled case can be obtained, while fulfilling all the RF emission requirements.

\subsection{Radio Link Level Evaluations}

In order to asses the performance of the 3MSK, several radio link level simulations have been evaluated under different conditions to compare the performance of the different variants of 3MSK in terms of uncoded bit-error-rate (BER) for a coherent receiver. These conditions include additive white Gaussian noise (AWGN) channel and TDL-E channel model defined in \cite{channelModels} with delay spread of 50 ns. Additionally, PN degradation is included in the simulations to show the robustness of the receiver processing against the effect of PN by using the models defined in \cite{PNModels}, where the phase tracking step of the 3MSK receiver is set to $\lambda = 0.05$. Finally, a comparison of block error rate (BLER) is evaluated for 3MSK by encoding the information with LDPC codes as defined in \cite{LDPCCodes} for the physical uplink shared channel (PUSCH) of 5G NR by using non-coherent receiver with the presence of PN.

\begin{figure}[t!]
	\centering
	\includegraphics[width=1.0\columnwidth]{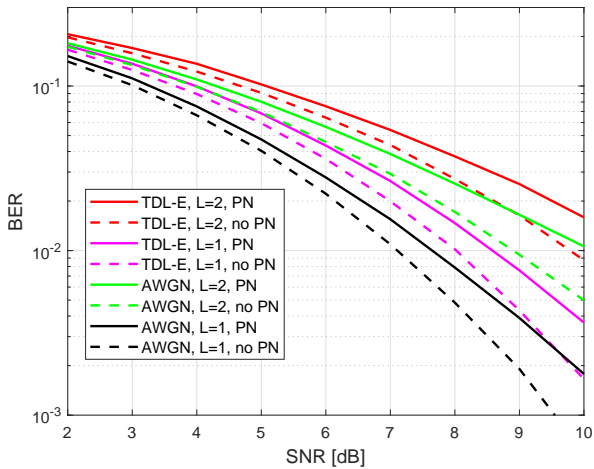}
	\caption{Link level results for uncoded BER in AWGN channel.}
	\label{fig:BERAWGN}
\end{figure}

Fig. \ref{fig:BERAWGN} shows the uncoded BER for AWGN channel between 3MSK with oversampling factors $L$ of 1 and 2, and QPSK, and with $a=0$ and $a=0.05$ for different EBW usage with full phase continuity for the narrow-band case of $K=12$. Generally, it was observed that phase continuity has minor effect on link performance and all BER results here are with full phase continuity. The receiver for the $L=2$ case when EBW is larger than 0\% directly discards the excess band, and only utilizes the in-band subcarriers for detection. It has to be noted that the extra gain in output power of the 3MSK compared to QPSK is not included in the illustration. It can be observed that the uncoded BER of 3MSK with $L=1$ is similar to that of QPSK, and that when oversampled transmission is included, there is a degradation of around 1 dB at the $10^{-1}$ BER point, and of around 2 dB at the $10^{-2}$ BER point. It could be expected that the degradation in required SNR for a given uncoded BER value can be compensated by the ability of generating more power out of the PA due to the lower PAPR of the signal. It is also worth to note that with $a=0.05$, the BER is lower than with $a=0$.

\begin{figure}[t!]
	\centering
	\vspace{-3mm}
	\includegraphics[width=1.0\columnwidth]{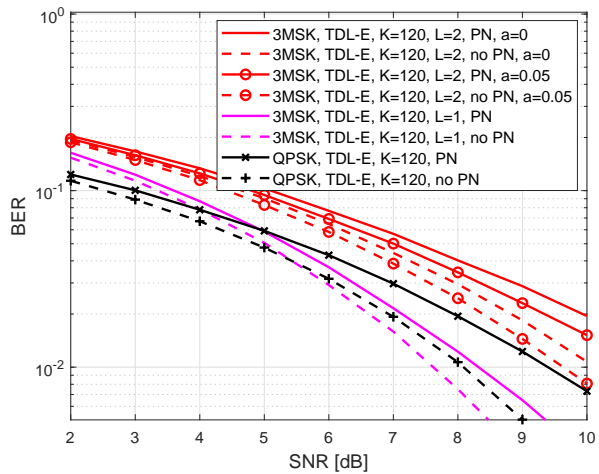}
	\vspace{-7mm}
	\caption{Link level results for TDL-E channel with and without PN for $L=1$ and $L=2$, $a=0$, and $a=0.05$.}
	\vspace{0mm}
	\label{fig:BERTDLE}
\end{figure}

Fig. \ref{fig:BERTDLE} shows the uncoded bit-error-rate (BER) for TDL-E channel for 3MSK with $L=1$ and $L=2$, and QPSK. For the 3MSK, the degradation when PN is included with $L=1$ is 0.5 dB at the $10^{-2}$ BER point, for both AWGN channel and TDL-E channel. The degradation for QPSK when PN is included is 1.3 dB, and 3MSK with $L=1$ performs 1 dB better than QPSK. Additionally, when $L=2$, 3MSK needs 1.2 dB higher SNR than QPSK to reach the $10^{-2}$ BER point, which is lower than the extra output power gain that could be obtained from the transmitter.  

\begin{figure}[t!]
	\centering
	\includegraphics[width=1.0\columnwidth]{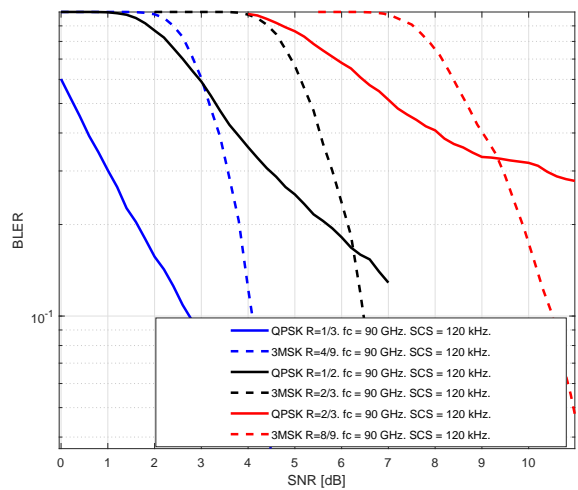}
	\vspace{-7mm}
	\caption{Coded Link level results for TDL-E channel with non-coherent receiver.}
	\vspace{-4mm}
	\label{fig:BLER}
\end{figure}

Finally, Fig. \ref{fig:BLER} shows the coded link level results for 3MSK with oversampling factor  $L=1$ and QPSK when non-coherent reception is assumed and PN is included. The coding rates of 3MSK and QPSK are paired in such a way that both transmissions have the same spectral efficiency. This means that the equivalent coding rate of 3MSK is increased by the factor of $2/1.5 = 1.33$ to ensure same spectral efficiency. For 3MSK, the phase error tracking step in the receiver, $\lambda$ is set to 0.05. It can be seen how for the 3MSK transmissions even with non-coherent receiver and severe PN degradation, 3MSK is able to reach the $10^{-1}$ BLER point, while QPSK presents an error floor when the coding rate is higher, proving the resilience of 3MSK to PN impairments.

\section{Conclusions}
\label{sec:conclusion}

In this article, a novel DFT-s-OFDM based waveform with 3-level CPM based modulation schemes, referred to as 3MSK, was proposed, shown to provide well-localized spectrum, low PAPR, and robustness against phase noise, while also facilitating non-coherent receiver processing and low processing complexity. It was observed that phase continuity between underlying CP-OFDM symbols helps to reduce the out-of-band emissions greatly. In terms of occupied bandwidth metric, 3MSK was found to exceed the performance of best $\pi/2$ BPSK DFT-s-OFDM reference schemes, also when a practical nonlinear power amplifier model is included. On the other hand, phase continuity between CP and main symbol, having clear effect on OOB emissions only in oversampled cases, has minor importance from the spectrum localization point of view. However, it helps to reduce PAPR in some scenarios, and most importantly, it helps to deal with phase uncertainty in the phase rotation based symbol phase continuity scheme, leading to improved link performance. Therefore, schemes with full phase continuity appear as the most interesting choice.

Oversampled signal generation and utilization of excess band was found to greatly improve the PAPR characteristics, reaching the PAPR of common $\pi/2$ BPSK schemes, at the cost of some loss in the link performance. One important advantage of the PAPR reduction can be seen in the results of maximum output power achievable from a realistic PA, where non-oversampled 3MSK can outperform QPSK by 3 dB, and with $L=2$, fully saturated PA can be used. The use of smoother 7-tap phase interpolation filter instead of linear interpolation provides worthwhile improvement both in PAPR and link performance. Finally, coded link level performance results were presented, showing that 3MSK without oversampling has similar performance to that of QPSK, and that in strong PN scenarios, 3MSK can be detected without the need of extra reference signals.

Overall, 3MSK was found to provide new and interesting tradeoffs between data rate and achievable transmission power with effective and low-cost power amplifiers, as an alternative to binary and 4-level transmission schemes. The views of future research include possibilities to reduce the gap in the link performance between basic and oversampled models through improved detection methods both for cases with and without excess band in the transmitter. In this paper, we considered only receivers which do not utilize the excess band, a scenario which has a benefit in terms of spectrum efficiency since the excess bands of different users/allocations may be overlapping. Also the utilization of the excess band on the receiver side is an interesting alternative to be considered in the future work.

\bibliographystyle{IEEEtran}
\bibliography{bibliography}

\begin{thebibliography}{10}
\providecommand{\url}[1]{#1}
\csname url@samestyle\endcsname
\providecommand{\newblock}{\relax}
\providecommand{\bibinfo}[2]{#2}
\providecommand{\BIBentrySTDinterwordspacing}{\spaceskip=0pt\relax}
\providecommand{\BIBentryALTinterwordstretchfactor}{4}
\providecommand{\BIBentryALTinterwordspacing}{\spaceskip=\fontdimen2\font plus
\BIBentryALTinterwordstretchfactor\fontdimen3\font minus
  \fontdimen4\font\relax}
\providecommand{\BIBforeignlanguage}[2]{{%
\expandafter\ifx\csname l@#1\endcsname\relax
\typeout{** WARNING: IEEEtran.bst: No hyphenation pattern has been}%
\typeout{** loaded for the language `#1'. Using the pattern for}%
\typeout{** the default language instead.}%
\else
\language=\csname l@#1\endcsname
\fi
#2}}
\providecommand{\BIBdecl}{\relax}
\BIBdecl

\bibitem{ChettriSurvey}
L.~Chettri and R.~Bera, ``{A Comprehensive Survey on Internet of Things (IoT)
  Toward 5G Wireless Systems},'' \emph{IEEE Internet of Things Journal},
  vol.~7, no.~1, pp. 16--32, 2020.

\bibitem{MoznimMTCB5G}
R.~Mozny, M.~Stusek, P.~Masek, K.~Mikhaylov, and J.~Hosek, ``{Unifying
  Multi-Radio Communication Technologies to Enable mMTC Applications in B5G
  Networks},'' in \emph{2020 2nd 6G Wireless Summit (6G SUMMIT)}, 2020, pp.
  1--5.

\bibitem{HoellerB5GLPWAN}
A.~Hoeller, J.~Sant'Ana, J.~Markkula, K.~Mikhaylov, R.~Souza, and H.~Alves,
  ``{Beyond 5G Low-Power Wide-Area Networks: A LoRaWAN Suitability Study},'' in
  \emph{2020 2nd 6G Wireless Summit (6G SUMMIT)}, 2020, pp. 1--5.

\bibitem{RazaOverview}
U.~Raza, P.~Kulkarni, and M.~Sooriyabandara, ``{Low Power Wide Area Networks:
  An Overview},'' \emph{IEEE Communications Surveys Tutorials}, vol.~19, no.~2,
  pp. 855--873, 2017.

\bibitem{AnnamalaiIoTPHY}
P.~Annamalai, J.~Bapat, and D.~Das, ``{Emerging Access Technologies and Open
  Challenges in 5G IoT: From Physical Layer Perspective},'' in \emph{2018 IEEE
  International Conference on Advanced Networks and Telecommunications Systems
  (ANTS)}, 2018, pp. 1--6.

\bibitem{FDSSpaper}
I.~Peruga~Nasarre, T.~Levanen, K.~Pajukoski, A.~Lehti, E.~Tiirola, and
  M.~Valkama, ``{Enhanced Uplink Coverage for 5G NR: Frequency-Domain Spectral
  Shaping With Spectral Extension},'' \emph{IEEE Open Journal of the
  Communications Society}, vol.~2, pp. 1188--1204, 2021.

\bibitem{nbIoTTutorial}
M.~Kanj, V.~Savaux, and M.~Le~Guen, ``{A Tutorial on NB-IoT Physical Layer
  Design},'' \emph{IEEE Communications Surveys Tutorials}, vol.~22, no.~4, pp.
  2408--2446, 2020.

\bibitem{LoRa}
``{LoRaWAN Specification}, {LoRa Alliance, Beaverton, OR, USA, 2016.
  [Online]},''
  \textit{\url{https://lora-alliance.org/lorawan-for-developers/}}, {Accessed:
  2021-09-10}.

\bibitem{DemikolMAC}
I.~Demirkol, C.~Ersoy, and F.~Alagoz, ``{MAC protocols for wireless sensor
  networks: a survey},'' \emph{IEEE Communications Magazine}, vol.~44, no.~4,
  pp. 115--121, 2006.

\bibitem{MSKPatent}
M.~L. Doelz and E.~H. Head, ``{Minimum-shift data communication system},''
  \emph{U.S. Patent 2 977 417}, 1961.

\bibitem{proakis_digital_2008}
J.~G. Proakis and M.~Salehi, \emph{Digital Communications}, 5th~ed.\hskip 1em
  plus 0.5em minus 0.4em\relax {McGraw}-Hill.

\bibitem{MSKPasupathy}
S.~Pasupathy, ``{Minimum shift keying: A spectrally efficient modulation},''
  \emph{IEEE Communications Magazine}, vol.~17, no.~4, pp. 14--22, 1979.

\bibitem{GMSK}
K.~Murota and K.~Hirade, ``{GMSK Modulation for Digital Mobile Radio
  Telephony},'' \emph{IEEE Transactions on Communications}, vol.~29, no.~7, pp.
  1044--1050, 1981.

\bibitem{PNcompIoT}
A.~Z. Mohammed, A.~K. Nain, J.~Bandaru, A.~Kumar, D.~S. Reddy, and
  R.~Pachamuthu, ``{A Residual Phase Noise Compensation Method for IEEE
  802.15.4 Compliant Dual-Mode Receiver for Diverse Low Power IoT
  Applications},'' \emph{IEEE Internet of Things Journal}, vol.~6, no.~2, pp.
  3437--3447, 2019.

\bibitem{IEEE802154}
``{IEEE Standard for Low-Rate Wireless Networks},'' \emph{IEEE Std
  802.15.4-2020 (Revision of IEEE Std 802.15.4-2015)}, pp. 1--800, 2020.

\bibitem{Kim_TVT2018}
J.~Kim, Y.~H. Yun, C.~Kim, and J.~H. Cho, ``{Minimization of PAPR for
  DFT-Spread OFDM With BPSK Symbols},'' \emph{IEEE Transactions on Vehicular
  Technology}, vol.~67, no.~12, pp. 11\,746--11\,758, 2018.

\bibitem{Myung_VTMAG2006}
H.~G. Myung, J.~Lim, and D.~J. Goodman, ``{Single carrier FDMA for uplink
  wireless transmission},'' \emph{IEEE Vehicular Technology Magazine}, vol.~1,
  no.~3, pp. 30--38, 2006.

\bibitem{Choi_OJ_COM_2021}
J.~Choi, J.~H. Cho, and J.~S. Lehnert, ``Continuous-phase modulation for
  {DFT}-spread localized {OFDM},'' \emph{IEEE Open Journal of the
  Communications Society}, vol.~2, pp. 1405--1418, 2021.

\bibitem{CPSK}
I.~Peruga~Nasarre, T.~Levanen, and M.~Valkama, ``{Constrained PSK:
  Energy-Efficient Modulation for Sub-THz Systems},'' in \emph{2020 IEEE
  International Conference on Communications Workshops (ICC Workshops)}, 2020,
  pp. 1--7.

\bibitem{Wylie_TCOM_2011}
M.~P. Wylie-Green, E.~Perrins, and T.~Svensson, ``Introduction to {CPM-SC-FDMA:
  A} novel multiple-access power-efficient transmission scheme,'' \emph{IEEE
  Transactions on Communications}, vol.~59, no.~7, pp. 1904--1915, 2011.

\bibitem{BCJR}
L.~Bahl, J.~Cocke, F.~Jelinek, and J.~Raviv, ``Optimal decoding of linear codes
  for minimizing symbol error rate (corresp.),'' \emph{IEEE Transactions on
  Information Theory}, vol.~20, no.~2, pp. 284--287, 1974.

\bibitem{viterbi}
A.~Viterbi, ``Error bounds for convolutional codes and an asymptotically
  optimum decoding algorithm,'' \emph{IEEE Transactions on Information Theory},
  vol.~13, no.~2, pp. 260--269, 1967.

\bibitem{sova}
J.~Hagenauer and P.~Hoeher, ``A viterbi algorithm with soft-decision outputs
  and its applications,'' in \emph{{1989 IEEE Global Telecommunications
  Conference and Exhibition 'Communications Technology for the 1990s and
  Beyond'}}, 1989, pp. 1680--1686 vol.3.

\bibitem{RezaiPN}
M.~Rezaei and K.~M.-P. Aghdam, ``{Phase Noise Reduction in Low Cost 24/77 GHz
  FMCW Sensors},'' in \emph{2018 18th Mediterranean Microwave Symposium (MMS)},
  2018, pp. 229--232.

\bibitem{MaffezzoniPNCorrelation}
P.~Maffezzoni, F.~Pepe, and A.~Bonfanti, ``{A Unified Method for the Analysis
  of Phase and Amplitude Noise in Electrical Oscillators},'' \emph{IEEE
  Transactions on Microwave Theory and Techniques}, vol.~61, no.~9, pp.
  3277--3284, 2013.

\bibitem{noauthor_3gpp_2019}
{{3GPP} {TS} 38.101-2 {V16}.2.0}, ``{User} {Equipment} ({UE}) radio
  transmission and reception; {Part} 2: {Range} 2 {Standalone}, {Tech}. {Spec}.
  {Group} {Radio} {Access} {Network}. {Rel}-16,'' Dec. 2019.

\bibitem{5628256}
Y.~{Wang} and Z.~{Luo}, ``{Optimized Iterative Clipping and Filtering for PAPR
  Reduction of OFDM Signals},'' \emph{IEEE Transactions on Communications},
  vol.~59, no.~1, pp. 33--37, 2011.

\bibitem{LDPCCodes}
{3GPP TS 38.212 V16.6.0 Release 16}, ``{5G; NR; Multiplexing and channel
  coding},'' Aug. 2021.

\bibitem{channelModels}
{{3GPP} {TR} 38.901 {V15}.1.0 }, ``{Study} on channel model for frequencies
  from 0.5 to 100 {GHz}; {Tech}. {Spec}. {Group} {Radio} {Access} {Network},''
  Sep. 2019.

\bibitem{PNModels}
{{3GPP} {TR} 38.803 {V2}.0.0 }, ``{Study on New Radio Access Technology; RF and
  co-existence aspects (Release 14)},'' March 2017.

\bibitem{Kuchi_ETT2012}
K.~Kuchi, ``Partial response {DFT}-precoded-{OFDM} modulation,'' \emph{Eur.
  Trans. Telecommun.}, vol.~23, no.~7, p. 632–645, 2012.

\bibitem{3GPP_low_par}
R.-. 3GPP TSG RAN WG1 Meeting Ad-Hoc Meeting~1901, ``Additional simulation
  results on low {PAPR RS},'' Jan. 2019.

\bibitem{80211adRapp}
``{IEEE P802.11 Wireless LANs, "TGad Evaluation Methodology", doc.: IEEE
  802.11-09/0296r16},'' Jan. 2009.

\end{thebibliography}

\end{document}